\documentclass[twocolumn,times]{aastex631}

\newcommand{\nickel}{$\rm ^{56}Ni$}
\newcommand{\cobalt}{$\rm ^{56}Co$}

\usepackage{graphicx}	%
\usepackage{amsmath}	%
\usepackage{amssymb}	%

\usepackage{threeparttable}
\usepackage{mathptmx}

\newcommand{\Msun}{\ensuremath{{M}_\odot}}

\shorttitle{SN\,2020wnt}
\shortauthors{Tinyanont et al.}
\graphicspath{{./}{}}

\begin{document}

\title{Supernova 2020wnt: An Atypical Superluminous Supernova with a Hidden Central Engine}

\author[0000-0002-1481-4676]{Samaporn Tinyanont}
\affiliation{Department of Astronomy and Astrophysics, University of California, Santa Cruz, CA 95064, USA}

\author[0000-0002-3352-7437]{Stan E. Woosley}
\affiliation{Department of Astronomy and Astrophysics, University of California, Santa Cruz, CA 95064, USA}

\author[0000-0002-5748-4558]{Kirsty Taggart}
\affiliation{Department of Astronomy and Astrophysics, University of California, Santa Cruz, CA 95064, USA}

\author[0000-0002-2445-5275]{Ryan J. Foley}
\affiliation{Department of Astronomy and Astrophysics, University of California, Santa Cruz, CA 95064, USA}

\author[0000-0003-1710-9339]{Lin Yan}
\affiliation{Division of Physics, Mathematics and Astronomy, California Institute of Technology, Pasadena, CA 91125, USA}

\author[0000-0001-9454-4639]{Ragnhild Lunnan}
\affiliation{The Oskar Klein Centre, Department of Astronomy, Stockholm University, AlbaNova, SE-106 91 Stockholm, Sweden}

\author[0000-0002-5680-4660]{Kyle W. Davis}
\affiliation{Department of Astronomy and Astrophysics, University of California, Santa Cruz, CA 95064, USA}

\author[0000-0002-5740-7747]{Charles D. Kilpatrick}
\affiliation{Center for Interdisciplinary Exploration and Research in Astrophysics (CIERA) and Department of Physics and Astronomy, Northwestern University, Evanston, IL, 60208, USA}

\author[0000-0003-2445-3891]{Matthew R. Siebert}
\affiliation{Space Telescope Science Institute, 3700 San Martin Drive, Baltimore, MD 21218, USA}

\author[0000-0001-6797-1889]{Steve Schulze}
\affiliation{The Oskar Klein Centre, Department of Physics, Stockholm University, AlbaNova, SE-106 91 Stockholm, Sweden}

\author[0000-0002-5221-7557]{Chris Ashall}
\affiliation{Department of Physics, Virginia Tech, Blacksburg, VA 24061, USA}

\author[0000-0002-1066-6098]{Ting-Wan Chen}
\affiliation{The Oskar Klein Centre, Department of Astronomy, Stockholm University, AlbaNova, SE-106 91 Stockholm, Sweden}

\author[0000-0002-8989-0542]{Kishalay De}
\affiliation{MIT-Kavli Institute for Astrophysics and Space Research, 77 Massachusetts Ave., Cambridge, MA 02139, USA}

\author[0000-0001-9494-179X]{Georgios Dimitriadis}
\affiliation{School of Physics, Trinity College Dublin, The University of Dublin, Dublin 2, Ireland}

\author[0000-0001-9584-2531]{Dillon Z. Dong}
\affiliation{Division of Physics, Mathematics and Astronomy, California Institute of Technology, Pasadena, CA 91125, USA}

\author[0000-0002-4223-103X]{Christoffer Fremling}
\affiliation{Division of Physics, Mathematics and Astronomy, California Institute of Technology, Pasadena, CA 91125, USA}

\author[0000-0003-4906-8447]{Alexander Gagliano}
\affiliation{Department of Astronomy, University of Illinois at Urbana-Champaign, 1002 W. Green St., IL 61801, USA}

\author[0000-0001-8738-6011]{Saurabh W. Jha}
\affiliation{Department of Physics and Astronomy, Rutgers, the State University of New Jersey, Piscataway, NJ, USA}

\author[0000-0002-6230-0151]{David O. Jones}
\affiliation{Gemini Observatory, NSF's NOIRLab, 670 N. A'ohoku Place, Hilo, Hawai'i, 96720, USA}

\author[0000-0002-5619-4938]{Mansi M. Kasliwal}
\affiliation{Division of Physics, Mathematics and Astronomy, California Institute of Technology, Pasadena, CA 91125, USA}

\author[0000-0003-2736-5977]{Hao-Yu Miao}
\affiliation{Graduate Institute of Astronomy, National Central University, 300 Zhongda Road, Zhongli, Taoyuan 32001, Taiwan}

\author[0000-0001-8415-6720]{Yen-Chen Pan}
\affiliation{Graduate Institute of Astronomy, National Central University, 300 Zhongda Road, Zhongli, Taoyuan 32001, Taiwan}

\author[0000-0001-8472-1996]{Daniel A. Perley}
\affiliation{Astrophysics Research Institute, Liverpool John Moores University, IC2, Liverpool Science Park, 146 Brownlow Hill, Liverpool L3 5RF, UK}

\author[0000-0002-7252-5485]{Vikram Ravi}
\affiliation{Division of Physics, Mathematics and Astronomy, California Institute of Technology, Pasadena, CA 91125, USA}

\author[0000-0002-7559-315X]{C\'{e}sar Rojas-Bravo}
\affiliation{Department of Astronomy and Astrophysics, University of California, Santa Cruz, CA 95064, USA}

\author[0000-0003-0466-3779]{Itai Sfaradi}
\affiliation{Racah Institute of Physics, The Hebrew University of Jerusalem, Jerusalem 91904, Israel}

\author[0000-0003-1546-6615]{Jesper Sollerman}
\affiliation{The Oskar Klein Centre, Department of Astronomy, Stockholm University, AlbaNova, SE-106 91 Stockholm, Sweden}

\author[0000-0001-5485-1570]{Vanessa Alarcon} 
\affiliation{Department of Astronomy and Astrophysics, University of California, Santa Cruz, CA 95064, USA}

\author[0000-0001-5281-731X]{Rodrigo Angulo} 
\affiliation{Department of Physics and Astronomy, The Johns Hopkins University, Baltimore, MD 21218, USA}

\author[0000-0001-8756-1262]{Karoli E.Clever} 
\affiliation{Department of Biomolecular Engineering and Bioinformatics, Baskin School of Engineering University of California, Santa Cruz, CA 95064, USA}

\author{Payton Crawford} 
\affiliation{Department of Astronomy and Astrophysics, University of California, Santa Cruz, CA 95064, USA}

\author[0000-0002-4653-9015]{Cirilla Couch} 
\affiliation{Department of Astronomy and Astrophysics, University of California, Santa Cruz, CA 95064, USA}

\author[0000-0003-4578-3216]{Srujan Dandu}
\affiliation{Department of Astronomy and Astrophysics, University of California, Santa Cruz, CA 95064, USA}

\author[0000-0002-5950-1702]{Atirath Dhara} 
\affiliation{Department of Astronomy and Astrophysics, University of California, Santa Cruz, CA 95064, USA}

\author{Jessica Johnson} 
\affiliation{Department of Astronomy and Astrophysics, University of California, Santa Cruz, CA 95064, USA}

\author[0000-0001-8483-9089]{Zhisen Lai}
\affiliation{Department of Astronomy and Astrophysics, University of California, Santa Cruz, CA 95064, USA}

\author{Carli Smith}
\affiliation{Ken and Mary Alice Lindquist Department of Nuclear Engineering, The Pennsylvania State University, University Park, PA 16802, USA}

\begin{abstract}
We present observations of a peculiar hydrogen- and helium-poor stripped-envelope (SE) supernova (SN) 2020wnt, primarily in the optical and near-infrared (near-IR). 
Its peak absolute bolometric magnitude of $-20.9$~mag and a rise time of 69~days are reminiscent of hydrogen-poor superluminous SNe (SLSNe~I), luminous transients potentially powered by spinning-down magnetars.
Before the main peak, there is a brief peak lasting $<$10~days post-explosion, likely caused by interaction with circumstellar medium (CSM) ejected $\sim$years before the SN explosion. %
The optical spectra near peak lack a hot continuum and \ion{O}{2} absorptions, which are signs of heating from a central engine; they quantitatively resemble those of radioactivity-powered H/He-poor Type Ic SESNe.
At $\sim$1~year after peak, nebular spectra reveal a blue pseudo-continuum and narrow \ion{O}{1} recombination lines associated with magnetar heating.
Radio observations rule out strong CSM interactions as the dominant energy source at +266~days post peak.
Near-IR observations at +200--300~day reveal carbon monoxide and dust formation, which causes a dramatic optical light curve dip. %
Pair-instability explosion models predict slow light curve and spectral features incompatible with observations.
SN\,2020wnt is best explained as a magnetar-powered core-collapse explosion of a 28~\Msun\ pre-SN star.
The explosion kinetic energy is significantly larger than the magnetar energy at peak, effectively concealing the magnetar-heated inner ejecta until well after peak.
SN\,2020wnt falls into a continuum between normal SNe~Ic and SLSNe~I and demonstrates that optical spectra at peak alone cannot rule out the presence of a central engine.
\end{abstract}

\keywords{Core-collapse supernovae(304) --- Massive stars(732)}

\section{Introduction} \label{sec:intro}

Massive stars, $\gtrsim 8 \, M_\odot$, conclude their evolution in many different flavors of core-collapse supernovae (CCSNe).
Stars that have lost their hydrogen envelope throughout their evolution, either via winds or binary interaction, produce a stripped-envelope (SE) SN, which shows little (Type IIb) to no sign of hydrogen (Type Ib) or helium (Type Ic) in their spectra around peak. 
Rapidly rotating stars may also undergo chemically homogeneous evolution (CHE), and lose hydrogen through nuclear burning.
In the past two decades, a subclass of CCSNe, superluminous (SL) SNe, with a total radiative energy about two orders of magnitude larger than that of a normal CCSN, has been discovered \citep[e.g.,][]{smith2007, quimby2007, ofek2007, barbary2009}; many of which aided by untargeted transient surveys \citep{quimby2011, chomiuk2011, nicholl2014}. 
(See reviews by \citealp{gal-yam2012, gal-yam2019a}.)
While the threshold for a SLSN was initially demarcated at $-21$~mag \citep{gal-yam2012}, later observations with a larger sample found that SLSNe form spectroscopic classes distinct from ordinary SNe \citep{inserra2013, nicholl2014, quimby2018, lunnan2018}; and that the two populations overlap in luminosity.
Similar to ordinary SNe, SLSNe show two distinct classes with (Type II) and without hydrogen (Type I). 
For hydrogen-rich SLSNe, interactions between the SN shock and circumstellar medium (CSM) can generally explain their luminosity and spectral properties \citep{chevalier2011}, though other power sources may contribute \citep{inserra2018}.
The power source for hydrogen-poor SLSNe is more debated \citep[e.g.,][]{moriya2018}.

Ordinary hydrogen-poor CCSNe (and also Type Ia SNe from explosions of white dwarfs in binary systems) are powered by the radioactive decay of \nickel\ and \cobalt. 
As such, the peak luminosity of these explosions is set by the total amount of \nickel\ produced, and the timescale of the light curve rise is set by the diffusion timescale of the ejecta, which is proportional to the total ejecta mass.
In most SLSNe~I, the peak luminosity is large but the light-curve rise is relatively short, such that the amount of \nickel\ required to power its peak equals or exceeds the total ejecta mass inferred from the rise time of the light curve \citep[][and references therein]{gal-yam2012}.

To explain the extra luminosity observed, a central engine, such as a spinning-down magnetar \citep{woosley2010, kasen2010} or an accreting black hole \citep{Mac01, dexter2013,  moriya2019}, is often invoked to add energy into the SN ejecta. 
In the magnetar model, a magnetized neutron star with with an initial period of $\sim 2 \, \rm ms$ and magnetic field of $\sim 10^{14} \, \rm G$ \citep{inserra2013, nicholl2017} is born after the core collapse. 
It is coupled to the SN ejecta and spins down depositing its rotational energy into the SN ejecta, with a time-dependent rate of $L\propto t^{-2}$ \citep{ostriker1971}, assuming a dipole radiation. 
In this model, it is crucial that the diffusion timescale of the SN ejecta and the magnetar spin-down timescale are roughly equal, allowing the magnetar heating to power a luminous peak \citep{inserra2013, nicholl2017}.
Given this requirement, the region of the ejecta affected by magnetar heating is readily visible during the optically thick phase of the SN.
As a result, spectroscopic signatures from the magnetar can be observed: most notably a hot blue continuum and W-shaped \ion{O}{2} absorption series in the blue part of optical spectra \citep[e.g.,][]{quimby2018}.
While the magnetar model is generally successful at explaining observations of SLSNe \citep[e.g.,][]{inserra2013, nicholl2017}, there remain some unexplained features. 
Many SLSNe~I show an early bump on a timescale of days after the explosion, likely from the shock cooling emission from an extended envelope above the stellar surface \citep{piro2015, nicholl_smartt2016, piro2021}.
Late-time light curves show bumps, which are unexpected from the power-law energy injection from a magnetar spin down and would require an additional power source, e.g., magnetar accretion and CSM interactions \citep{nicholl2016a, inserra2017, hosseinzadeh2021, chen2022b}. 
Recent population studies of SLSNe~I observed by the Zwicky Transient Facility (ZTF) have found that about 20\% of SLSNe~I require earlier CSM interactions to fit their light curves \citep{chen2022a, chen2022b}.
A small number of SLSNe~I indeed show spectroscopic signatures of interactions with H-rich CSM at late times \citep{yan2015, yan2017b, chen2018}.

Some spectroscopic SLSNe evolve slowly enough such that radioactivity could explain their light curves. 
If these SLSNe are truly powered by radioactivity, the \nickel\ mass required (few \Msun) far exceeds what the standard neutrino-driven core collapse mechanism can produce, which is about 0.2 $M_\odot$ observed in SNe II \citep{rodriguez2021} and SESNe \citep{lyman2016, sravan2020}.
The only theoretically well-established physical mechanism that is able to produce such amount of \nickel\ is the pair-instability (PI) process \citep[e.g.,][]{rakavy1967, barkat1967, heger2002, kasen2011}. 

Pair-Instability (PI) SNe happen in extremely massive stars with the initial mass between 140 and 260 \Msun \citep{kasen2011}.
At the onset of carbon burning, the conditions in the core allow for rapid electron/positron pair production, which removes the radiation pressure support. %
As a result, the core collapses igniting oxygen explosively, disrupting the star.
The resulting PISNe are predicted to have a slow photometric evolution ($\gtrsim$100 d rise time) due to the large ejecta mass and distinct spectroscopic signatures due to its thermonuclear nature and low velocity.
Thus far, a few slowly evolving SLSNe have been proposed as PISN candidates (e.g., SN\,2007bi, \citealp{gal-yam2009}, but see \citealp{young2010}).
There are also a few other events evolving on a similar timescale (e.g., PTF12dam, \citealp{nicholl2013, chen2015,vreeswijk2017}; PS1-14bj, \citealp{lunnan2016}).
However, none of them have spectroscopic signatures consistent with PISN models \citep[e.g.,][]{dessart2012, jerkstrand2016, jerkstrand2017, mazzali2019, moriya2019}.
In general, these PISN candidates' spectra are too blue and the lines observed are too broad, and the nebular spectra show different abundance pattern from what is expected from a PISN. 
Because of the spectroscopic discrepancy, some argued that these events could be core-collapse explosions, with novel explosion mechanisms, of very massive stripped stars\citep[e.g.,][]{mazzali2019, moriya2019}.

Peculiarities also exist in the spectroscopic SESN population. 
For instance, there are spectroscopic SNe~Ib/c with late-time evolution inconsistent with a single radioactive power source. 
SN\,2010mb \citep{ben-ami2014} shows extra luminosity, blue pseudo-continuum, and narrow [\ion{O}{1}] emission at late times, attributed to interactions with $\sim$3~$M_\odot$ of H-poor CSM. 
Other events like iPTF15dtg have a relatively long rise time and high peak luminosity, indicative of a large \nickel\ mass; but late-time observations show a power-law tail much better fit by a magnetar model \citep{taddia2016, taddia2019}.
Some multi-peak events like SN\,2019stc may require radioactivity, a magnetar, and CSM interactions to explain \citep{gomez2021}.
The diversity in the observed properties of H/He-poor SNe is a manifestation of the different \nickel\ mixing, stripping mechanisms, CSM interactions, and the degree at which the new-born neutron star affects the resulting SN \citep[e.g.,][]{afsariardchi2021,sollerman2022, gomez2022}. 
Peculiar events that probe this vast range of SESN properties are still routinely being discovered.

Here we present observations of SN\,2020wnt, a H- and He-poor SN with distinct photometric and spectroscopic properties.
Its light curve shape resembles that of SESNe, showing a relatively symmetric peak falling onto an exponential decline tail.
However, the peak luminosity is much larger and the rise time is much longer than those of SESNe. 
Its spectroscopic evolution closely resembles that of SNe Ic up to about 1 year post peak with many marked differences compared to SLSNe.
Late-time optical spectra, however, reveal features similar to magnetar-powered SLSNe.
Late-time near-infrared (IR) observations reveal the formation of carbon monoxide (CO) and dust, similar to what is observed in SESNe and relatively novel for SLSNe. 
We recognize that while preparing this paper, \cite{gutierrez2022} posted their paper on the same supernova on the arXiv, presenting some similar observations and analysis; this work should be treated as an independent analysis on a largely independent data set (with only shared public ATLAS and ZTF photometry). 
In \textsection\ref{sec:obs} we summarize the discovery and follow-up observations of SN\,2020wnt.
In \textsection\ref{sec:analysis} we analyze photometric data and compute explosion properties.
In \textsection\ref{sec:spec_evo} we discuss the optical to near-IR spectroscopic evolution of SN\,2020wnt, quantitatively comparing it to SESNe and SLSNe.
In \textsection\ref{sec:model_comparison} we compare our bolometric luminosity to various models to discern the nature of SN\,2020wnt.
In \textsection\ref{sec:nebular_spec_modeling} we compare our nebular spectra to model spectra to measure the properties of the ejecta.
In \textsection\ref{sec:host} we discuss the stellar mass, star-formation rate, and metallicity of the host galaxy.
We provide a discussion and conclusion in \textsection\ref{sec:conclusion}.

\begin{figure*}[!ht]
    \centering
    \includegraphics[width = 0.8\linewidth]{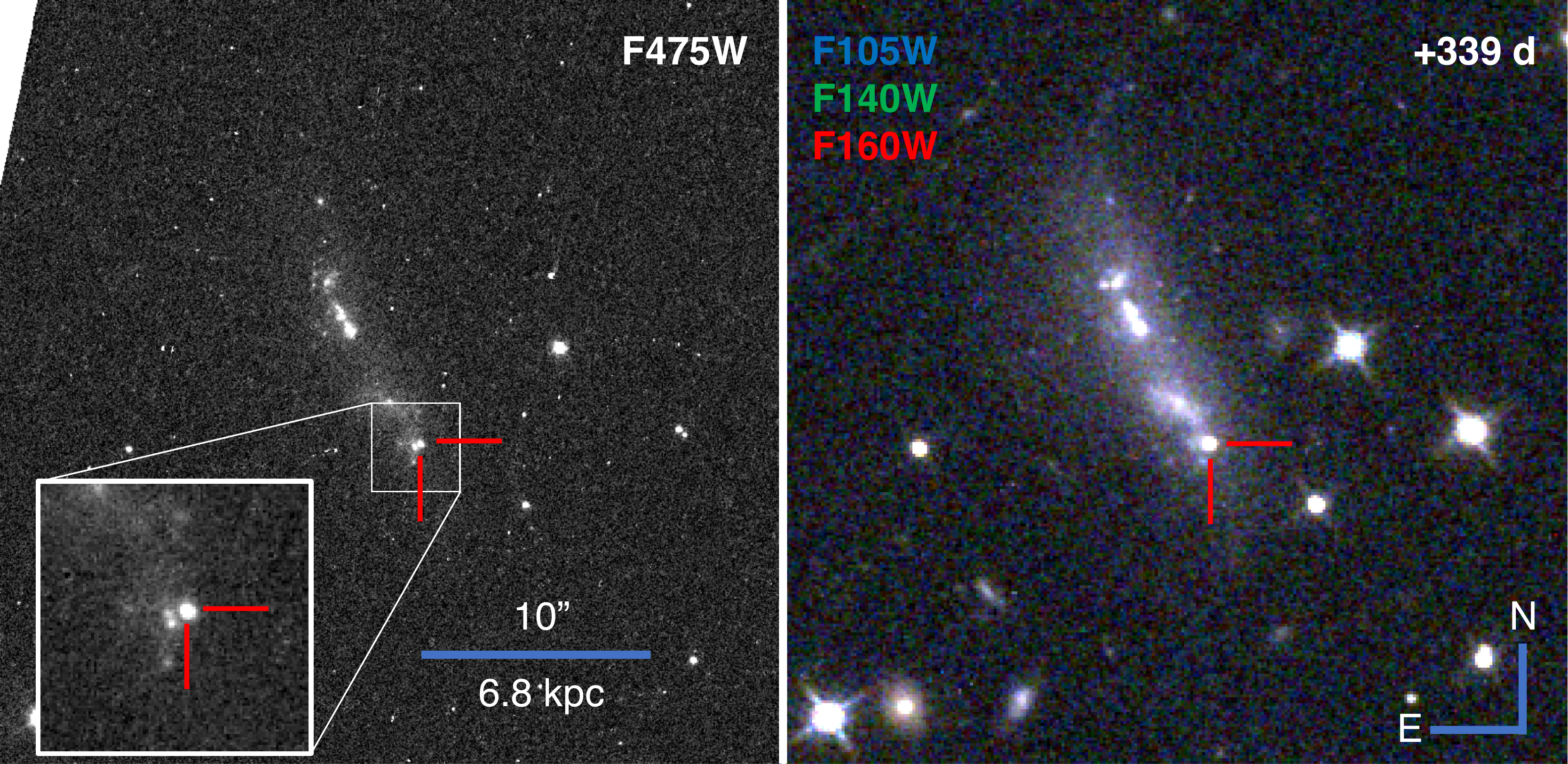}
    \caption{Images of SN\,2020wnt at 339 d post peak taken by \textit{HST}/WFC3 with the F475W filter (left, greyscale); and the F105W, F140W, and F160W filters (right, false color with the respective filters corresponding to blue, green, and red). The SN is marked with red crosses and the angular scale is notated. North is up and East is to the left. The inset of the left panel shows a zoomed-in image of the SN location. There are at least two neighboring sources to the SN, which are likely \ion{H}{2} regions in the host galaxy.}
    \label{fig:images}
\end{figure*}

\section{Observations}\label{sec:obs}

\subsection{Supernova Discovery and Classification}
SN\,2020wnt (ZTF20acjeflr) was discovered by the Zwicky Transient Facility \citep[ZTF;][]{bellm2019, graham2019, masci2019} through the event broker Automatic Learning for the Rapid Classification of Events \citep[ALeRCE;][]{forster2020} on 2020 October 14 UT \citep{forster2020b} (UT dates used hereafter).
The discovery magnitude was $g = 19.7$. 
We decided to start following up this SN based on public light curves gathered by \texttt{YSE-PZ}, our Target and Observation Management System \citep{coulter2022}.
We classified SN\,2020wnt as a Type~I SN on 2020 November 16 using an optical spectrum obtained with the Kast spectrograph on the 3-meter Shane Telescope at Lick Observatory \citep{tinyanont2020}.
All subsequent photometric and spectroscopic observations are also organized using \texttt{YSE-PZ}.
The classification spectrum contained a narrow H$\alpha$ emission from the host galaxy, putting the SN at $z=0.0323 \pm 0.0001$.
The corresponding luminosity distance is 141.8 Mpc, assuming a standard $\rm \Lambda CDM$ cosmology with $H_0 = 70 \ \rm km \, s^{-1} \, Mpc^{-1}$, $\Omega_M = 0.3$ and $\Omega_\Lambda = 0.7$.
At the time of classification, the transient had been brightening for a month.
The Galactic extinction along the line of sight toward SN\,2020wnt is $E(B-V) = 0.42$~mag \citep{schlafly2011}.
We use this value for extinction correction throughout the paper, assuming $R_V = 3.1$ \citep{cardelli1989} and the extinction law of \citep{fitzpatrick1999}. 
We assume that the host extinction is negligible owing to the lack of the \ion{Na}{1} D absorption at the host redshift. 

\subsection{Photometry}
We obtained public forced photometry of SN\,2020wnt from ZTF in the $g$ and $r$ bands, and from the Asteroid Terrestrial-impact Last Alert System (ATLAS; \citealp{tonry2018, smith2020}) in the cyan and orange bands.
The SN was observed by these public surveys at a cadence of a few days.

These regularly-scheduled photometry were supplemented by observations in the \textit{griz} bands from the 2-m Liverpool Telescope (LT) on La Palma; \textit{BVgriz} bands from the Lulin One-meter Telescope (LOT) at Lulin Observatory in Taiwan; and \textit{BVri} bands from the 1-m Nickel telescope at Lick Observatory in California. 
These observations were reduced using a standard optical imaging procedures: bias and flat-field correction and photometric calibration using stars observed in the same field of view. 

We analyzed the \textit{griz} imaging from the LOT and the \textit{ri} imaging from the Nickel telescope using image frames from the Pan-STARRS 3$\pi$ survey \citep{Flewelling16} as templates.  After registering and estimating a zero point for each LOT image in {\tt photpipe} \citep{Rest05}, we performed digital image subtraction between each LOT and Pan-STARRS image using {\tt hotpants} \citep{Becker15}.  We then performed forced photometry at the location of SN\,2020wnt using a custom version of {\tt DoPhot} \citep{Schechter93}.

We obtained ground-based near-IR imaging of SN\,2020wnt on 2021 August 19 using the slit-viewing camera of the SpeX spectrograph \citep{rayner2003} on the NASA InfraRed Telescope Facility (IRTF) in the $J$, $H$, and $K$ bands.
The data were reduced using a custom \texttt{python} script that constructed the sky flat image from the dithered observations, performed flat fielding and background subtraction, then shifted and coadded observations in each band. 
We obtained another epoch of near-IR photometry on 2021 December 10 using the Near-InfraRed Imager (NIRI) on Gemini North in the $J$, $H$, and $Ks$ bands as part of the fast turnaround program GN-2021B-FT-109 (PI Tinyanont).  
We used \texttt{DRAGONS v. 3.0.1} to reduce NIRI images; the steps were similar to the script we used to reduce IRTF images. 
Photometric calibration in both cases were obtained using 2MASS \citep{milligan1996,skrutskie2006} stars in the field of view.

In addition to ground-based observations, we observed SN\,2020wnt with the Ultra-Violet and Optical Telescope \citep[UVOT][]{roming2005} aboard the \textit{Neil Gehrels Swift Observatory} \citep{gehrels2004} at 46, 98, 118, 138, and 158 days post-discovery in the $U$, $B$, $V$, $UVW1$, and $UVW2$ bands.
The data were processed by the standard data reduction pipeline and obtained via the HEASARC archive. 
We did not detect the SN in the UV bands.
Aperture photometry was obtained using a 3$''$ radius aperture centered on the SN with local background subtraction. 

SN\,2020wnt was observed during the ongoing NEOWISE all-sky mid-IR survey in the $W1$ ($3.4$\,$\mu$m) and $W2$ ($4.5$\,$\mu$m) bands \citep{Wright2010, Mainzer2014}. We retrieved time-resolved coadded images of the field created as part of the unWISE project \citep{Lang2014, Meisner2018}. To remove contamination from the host galaxies, we used a custom code \citep{De2019} based on the ZOGY algorithm \citep{Zackay2016} to perform image subtraction on the NEOWISE images using the full-depth coadds of the WISE and NEOWISE mission (obtained during 2010-2014) as reference images. Photometric measurements were obtained by performing forced PSF photometry at the transient position on the subtracted WISE images until the epoch of the last NEOWISE data release (data acquired until December 2021).

\begin{figure*}
    \centering
    \includegraphics[width = 0.9\linewidth]{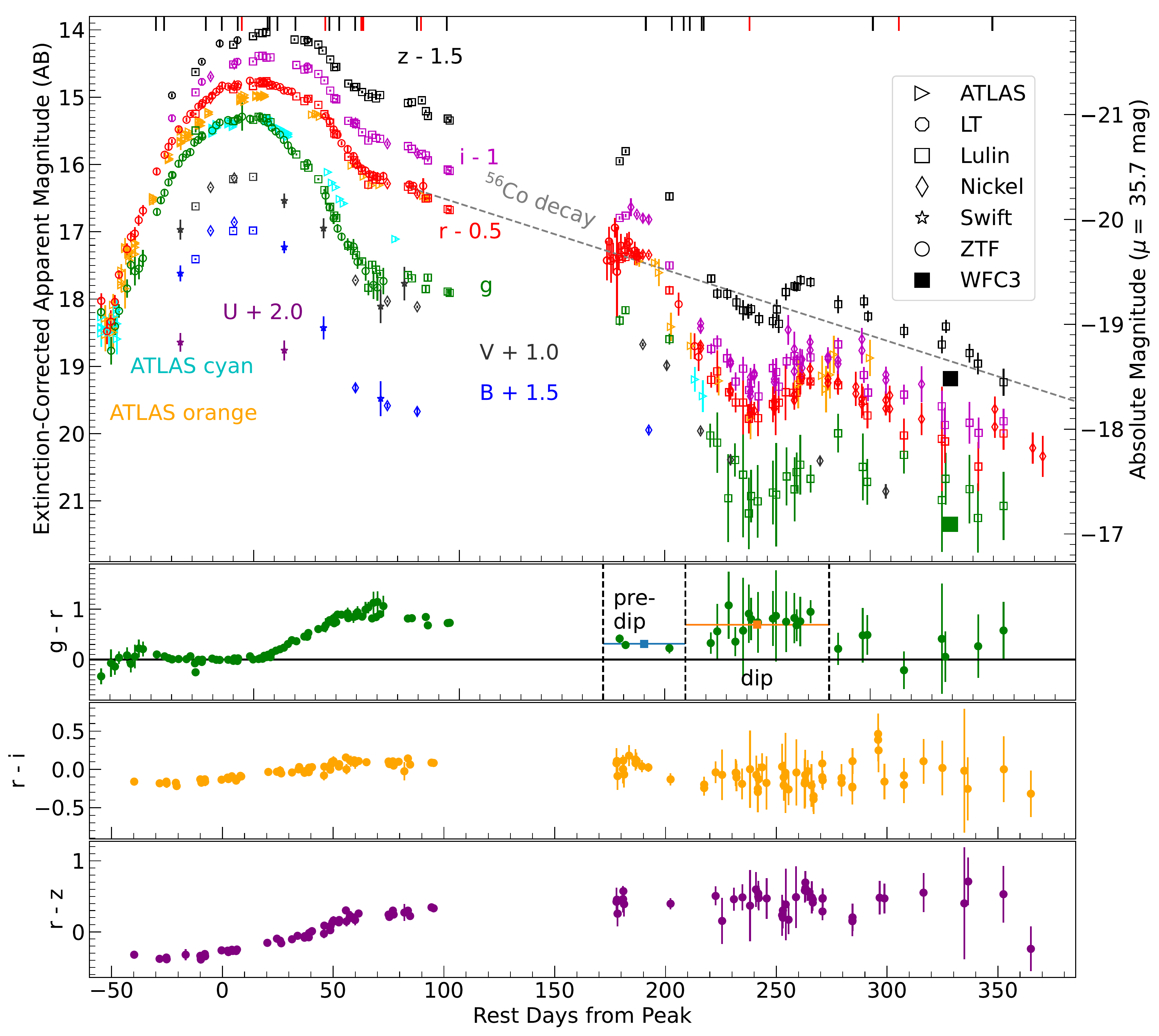}
    \caption{ \textbf{Top:} Multi-band photometry of SN\,2020wnt. All magnitudes are in the AB system (including the \textit{UBV} bands). Different symbol shapes denote the telescope/instrument used. The two WFC3 photometry points plotted are in F475W (in green) and F110W (in black). The photometry has been corrected for Galactic extinction using $E(B-V) = 0.42$~mag and $R_V = 3.1$. Different bands are offset by the specified amount to improve visibility. The absolute magnitude is computed with a distance modulus of $\mu =  35.74$~mag. Black (red) ticks on top of the figure indicate the epochs with optical (near-IR) spectroscopy. The 0.0098 $\rm~mag \, d^{-1}$ decline rate expected from the \cobalt\ decay is shown for comparison. 
    \textbf{Bottom:} The evolution of $g-r$, $r-i$, and $r-z$ colors of SN\,2020wnt.
    For the $g-r$ color, We mark the two regions around the light curve dip and plot the average color pre-dip (170--210 d) and during dip (210--280 d). 
    }
    \label{fig:photometry}
\end{figure*}

Lastly, we observed SN\,2020wnt with the Wide-Field Camera 3 (WFC3) on board the \textit{Hubble Space Telescope} (\textit{HST}) in three epochs on 2021 December 15, 2022 January 25, and 2022 August 26 (GO--16768, PI Tinyanont; SNAP--16691, PI Foley). 
The first epoch consists of imaging with the F275W and F475W filters in the UVIS channel and the F105W, F140W, and F160W filters in the IR channel. 
The second epoch consists of shorter exposures with the F105W and F140W filters, along with spectroscopy with the G102 and G141 grisms. 
We will discuss the grism spectroscopy below. 
Figure~\ref{fig:images} shows images of SN\,2020wnt in different bands, while Fig.~\ref{fig:photometry} shows the light curves of SN\,2020wnt. 

\subsection{VLA Radio observations}
\label{subsec:radio}
SN\,2020wnt was observed with the Karl G. Jansky Very Large Array (VLA) under the Director's Discretionary Time program (PI: Yan, program ID: 21B-161). 
The observations were carried out on 2021 September 24 UT using the B-configuration for C and Ku band, each with a total of 42\,minutes. The data were reduced using the standard procedures provided by the VLA software package Common Astronomy Software Applications (CASA).
SN\,2020wnt is not detected in the cleaned maps, which reach $1\sigma$ root-mean-squared of $5.4$ and $6.5 \ \mu$Jy at C and Ku band, respectively. 

\subsection{Optical Spectroscopy}
We obtained optical spectra of SN\,2020wnt primarily with the Kast spectrograph \citep{KAST} on the 3-m Shane Telescope at Lick Observatory and the Low-Resolution Imaging Spectrograph (LRIS; \citealp{oke1995}) on the Keck I telescope.
The log of spectroscopic observations (both optical and near-IR in the next section) is provided in Table~\ref{tab:spec_log}. 
Note that the last spectrum taken on 2022 September 22, 611 days post peak light contains no discernible SN light; we use this spectrum with the longest exposure time for host analysis. 
The spectra were reduced using a custom data reduction pipeline based on the Image Reduction and Analysis Facility (IRAF) \citep{tody1986}.\footnote{The pipeline is publicly accessible at \url{https://github.com/msiebert1/UCSC_spectral_pipeline}.}
The pipeline performed flat field corrections using observations of a flat field lamp.
The wavelength solution was derived using observations of arc lamps.
The instrument response function was derived using observations of spectroscopic standard stars. 
The 2D spectra were extracted using the optimal extraction algorithm \citep{horne1986}. 

We supplement this dataset with spectra from the Alhambra Faint Object Spectrograph and Camera (ALFOSC) on the 2.56-meter Nordic Optical Telescope (NOT) on La Palma and the Double Spectrograph (DBSP) at the 200-inch Hale Telescope at Palomar Observatory \citep{oke1982}.
These data were reduced using similar steps as described above.

\subsection{Infrared Spectroscopy}
We obtained near-IR (1--2.5 $\mu$m) spectroscopy of SN\,2020wnt at eight epochs, spanning $-5$ to 378 days from peak. 
We used the Near-Infrared Echellette Spectrometer (NIRES) on the Keck II telescope; TripleSpec \citep{herter2008} on the 200-inch Telescope at Palomar Observatory; SpeX IRTF; and the Gemini Near InfraRed Spectrograph \citep[GNIRS;][]{elias2006a,elias2006b} on the Gemini North telescope. 
NIRES and TripleSpec shared the same design and both provided a simultaneous wavelength coverage from 0.95 to 2.45~$\mu$m at $R \sim 2700$. 
They had 0\farcs55 and 1\farcs0 wide slits, respectively.
The first epoch of SpeX observation was obtained as part of the program 2021A044 (PI Tinyanont), using the short cross-dispersed (SXD) mode paired with the 0\farcs8 wide slit, providing simultaneous wavelength coverage from 0.7 to 2.55~$\mu$m at a resolving power of $R\sim$750.
The second epoch, as part of 2021B058 (PI Tinyanont), uses the low-resolution prism with the 0.8$''$ slit with a resolving power of $R\sim$70.
GNIRS data were obtained as part of the fast-turnaround programs GN-2021A-FT-104 and GN-2021B-FT-106 (PI Tinyanont). 
We used the cross-dispersed mode with the short camera paired with the 32 l/mm grating and the 0\farcs45 wide slit, providing simultaneous wavelength coverage from 0.8 to 2.5 $\mu$m at a resolving power of $R\sim$1100.
At all epochs, we observed an A0V star for telluric corrections immediately before or after the science observations. 
We reduced the NIRES, TripleSpec, and SpeX data using \texttt{spextool} \citep{cushing2004}; and the GNIRS data using Gemini IRAF data reduction package.  
We performed telluric corrections for all ground-based data using \texttt{xtellcor} \citep{vacca2003}.

The last epoch of near-IR spectroscopy was performed using \textit{HST}/WFC3 IR grisms G102 and G140. 
We obtained direct images in the F105W and F141W filters to provide the wavelength solution for the grism data based on the image position.
We performed data reduction and spectral extraction using the package \texttt{HSTaXe}, following the official cookbook.\footnote{Accessed via \url{https://github.com/npirzkal/aXe_WFC3_Cookbook}}

All spectroscopic data will be made available through WISeREP\footnote{\url{https://www.wiserep.org/}} \citep{yaron2012}.

\begin{figure*}[t]
    \centering
    \includegraphics[width=\textwidth]{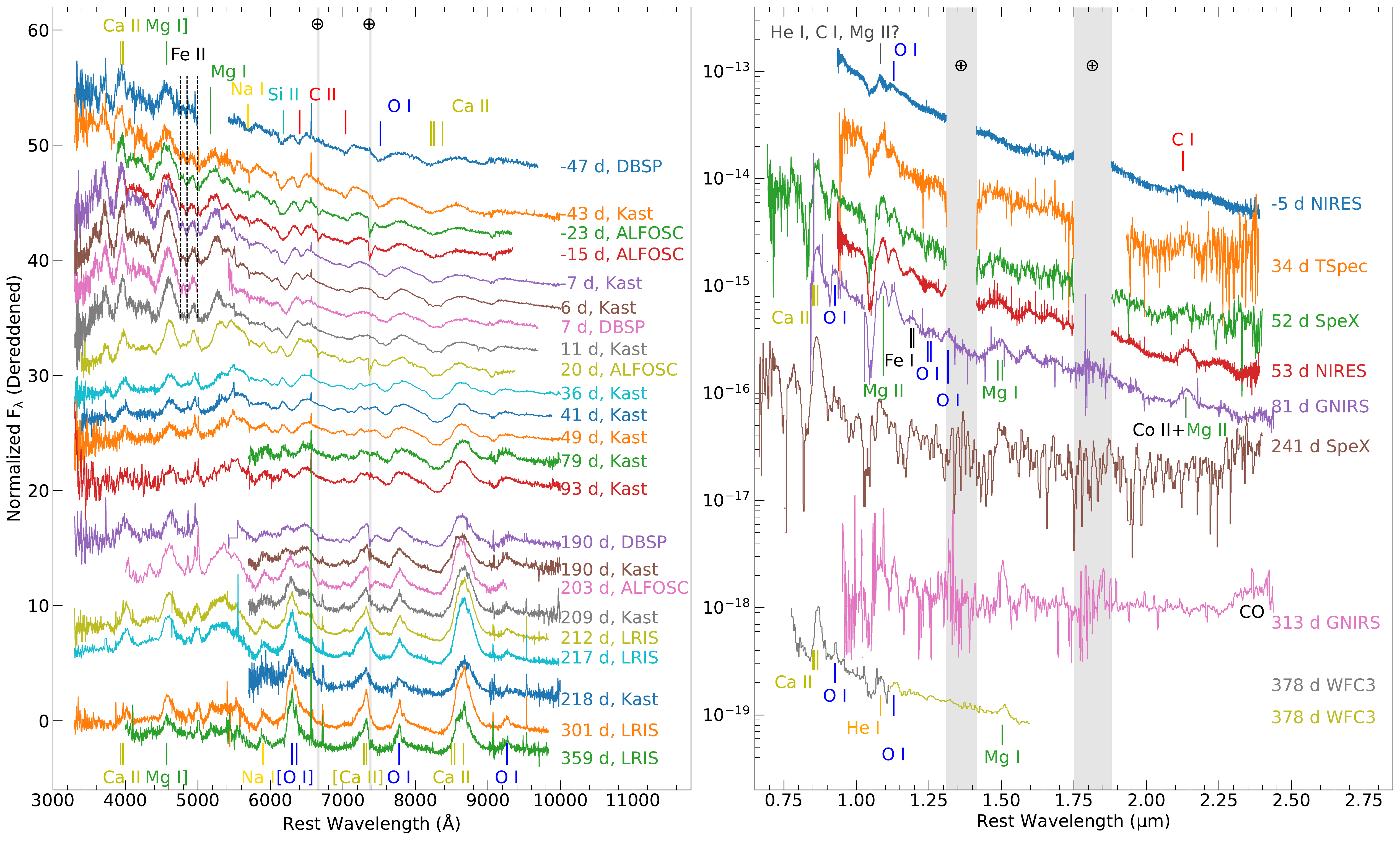} 
    \caption{Optical (left) and near-IR (right) spectra of SN\,2020wnt. Epoch of observation with respect to peak and instrument used for each observation is notated. The near-IR spectra are plotted with a log scale on the \textit{y} axis to improve visibility. Spectra are also binned at varying resolutions for the same purpose.
    Prominent spectral features are marked. 
    At early time, the \ion{Ca}{2} H\&K, \ion{Mg}{1}], and \ion{Mg}{1} lines are marked at rest; the \ion{Fe}{2} lines are marked at $-10${,}$000 \rm \ km \, s^{-1}$; and the rest of the lines are marked at $-8${,}$000 \rm \  km \, s^{-1}$.
    The emission lines at late times and all near-IR lines are marked at rest velocity. 
    Telluric bands are also marked.  }
    \label{fig:opt_spec}
\end{figure*}

\section{Light Curve Analysis}\label{sec:analysis}

\subsection{Bolometric Luminosity}\label{sec:bolo}
We compute the bolometric luminosity of SN\,2020wnt to capture the evolution of its radiative output using the following steps. 
We first interpolate the photometric data in different bands onto a common time grid using the Gaussian process regression package \texttt{george} \citep{ambikasaran2015} to perform the interpolation.
We use the exponential squared kernel with a timescale of 500 d (we found that the result does not depend sensitively on the timescale). 
The interpolation is performed in the flux space. 
We do not use data extrapolated for more than two days from actual observations in further analyses.

With the interpolated light curve, we compute the bolometric light curve using \texttt{SuperBol} \citep{nicholl2018}, which fits a black body model reddened by the Milky Way dust extinction law ($R_V = 3.1$; \citealp{cardelli1989}), with $E(B-V) = 0.42$~mag. 
We do not use any further interpolation in \texttt{SuperBol}. 
Most of our photometric follow-up observations begin after our spectroscopic classification at 43 d pre-peak; prior to that all photometric data come from ZTF and ATLAS. 
To avoid extrapolation, we run \texttt{SuperBol} for early-time ZTF and ATLAS photometry and for late-time photometry with more photometric bands separately. 
There is no jump in the resulting bolometric light curve. 
We find that suppressing flux at wavelengths shorter than 3000 \AA\, to mimic the effects of line blanketing does not affect the results because there is already little flux at those wavelengths. 
For each epoch, \texttt{SuperBol} outputs the fitted black body temperature and radius. 
The bolometric luminosity is then the sum of the observed luminosity in all bands and the unobserved luminosity inferred from the black body fit.

We compute two peak epochs: in the $r$ band and bolometric. 
In the $r$ band, we find the peak at MJD $=$ 59213, 75 d after the explosion. 
We use this peak epoch as a reference throughout the paper since it is less sensitive to uncertainties related to the bolometric fitting. 
From the bolometric light curve, we find the bolometric peak epoch and luminosity to be $69 \pm 2$ days post explosion in the rest frame and $L_{\rm bol, peak} = (6.8\pm 0.3) \times 10^{43} \rm \, erg\, s^{-1}$ by fitting low-order polynomial to the light curve around peak. 
This peak time is used to compute the SN explosion properties.
The uncertainties are derived from a Markov-Chain Monte Carlo fit performed using \texttt{emcee} \citep{foreman-mackey2013}.
The total emitted energy as of our last epoch of observation is $\sim 5 \times 10^{50} \, \rm erg$. 

\begin{figure*}
    \centering
    \includegraphics[width=0.49\linewidth]{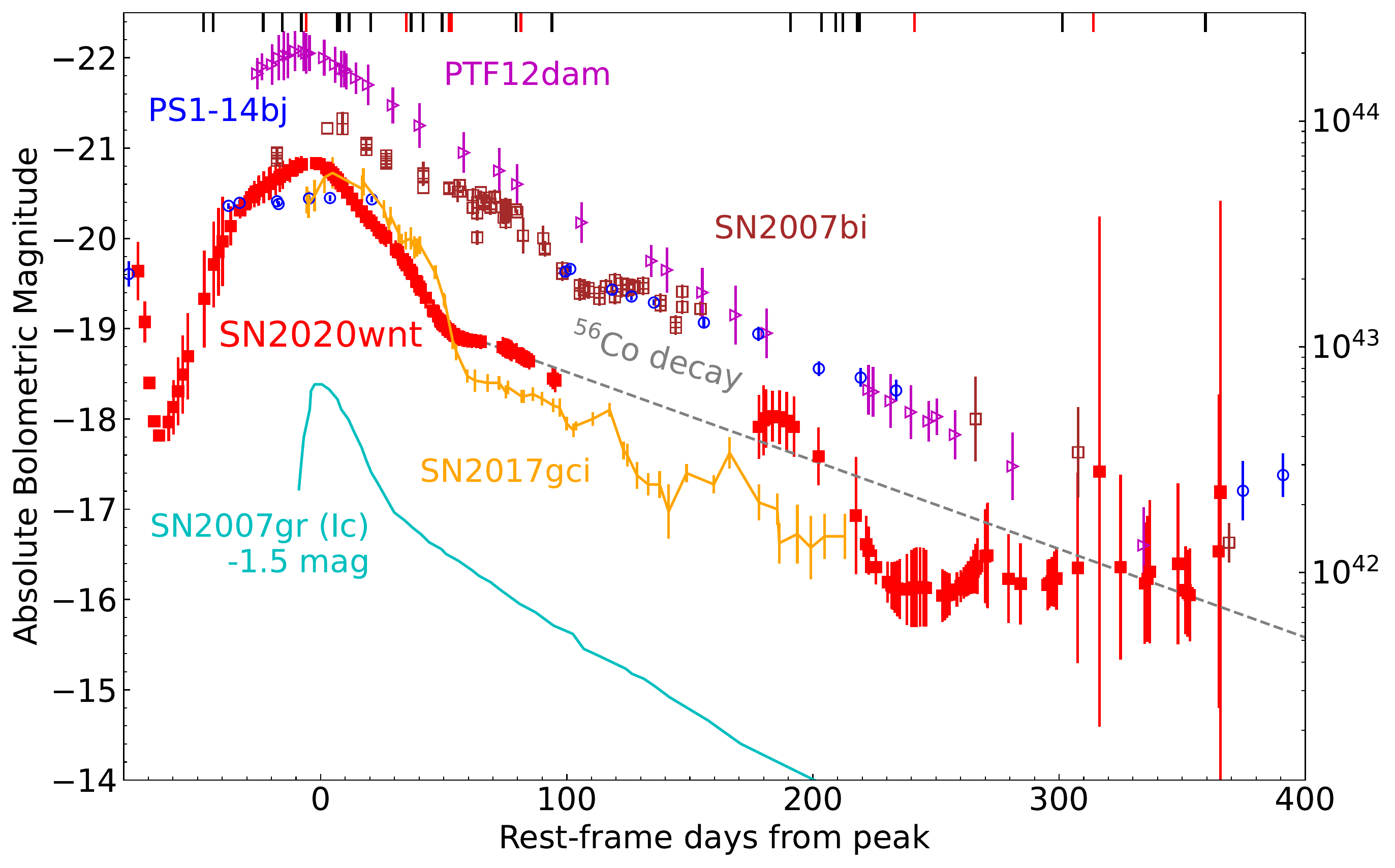} 
    \includegraphics[width=0.49\linewidth]{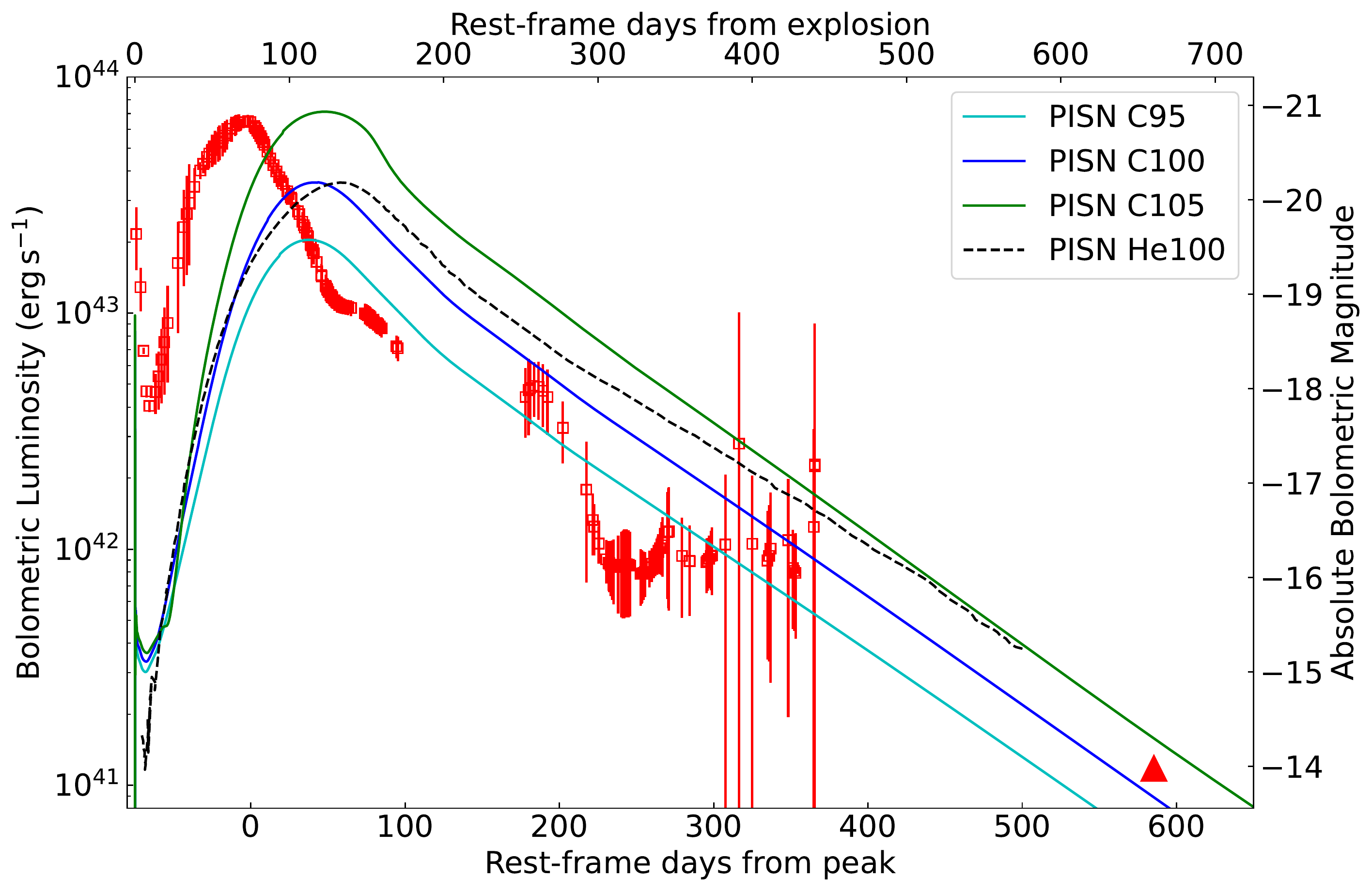}
    \caption{\textbf{Left: }Bolometric light curve of SN\,2020wnt compared with slowly evolving SLSNe (SN\,2007bi, \protect\citealp{gal-yam2009}; PTF12dam, \protect\citealp{vreeswijk2017}; PS1-14bj, \protect\citealp{lunnan2016}; and SN\,2017gci, \protect\citealp{fiore2021}) and a Type Ic (SN\,2007gr \protect\citealp{hunter2009}). The knee in the light curve at 50 d post peak and the late-time decline rate make SN\,2017gci the closest analog of SN\,2020wnt. 
    Black (red) ticks on top of the figure indicate the epochs with optical (near-IR) spectroscopy. 
    \textbf{Right: }Bolometric light curve of SN\,2020wnt compared with PISN models discussed in \textsection\ref{sec:model_comparison}.
    The red triangle denotes the lower limit of the bolometric luminosity obtained from the \textit{HST} observations of SN\,2020wnt 585 d post peak. 
    PISN models with 100~$M_\odot$ He star from \protect\citet{kasen2011} along with three new H/He-poor models with 95, 100, and 105 \Msun\,are plotted with dotted lines. 
    The comparison shows that PISN models rise too slowly to explain SN\,2020wnt's peak luminosity.
    Comparison with magnetar models are shown in Fig.~\ref{fig:LC_magnetar}.
    }
    \label{fig:bolometric_LC}
\end{figure*}

\begin{figure*}
    \centering
    \includegraphics[width=\linewidth]{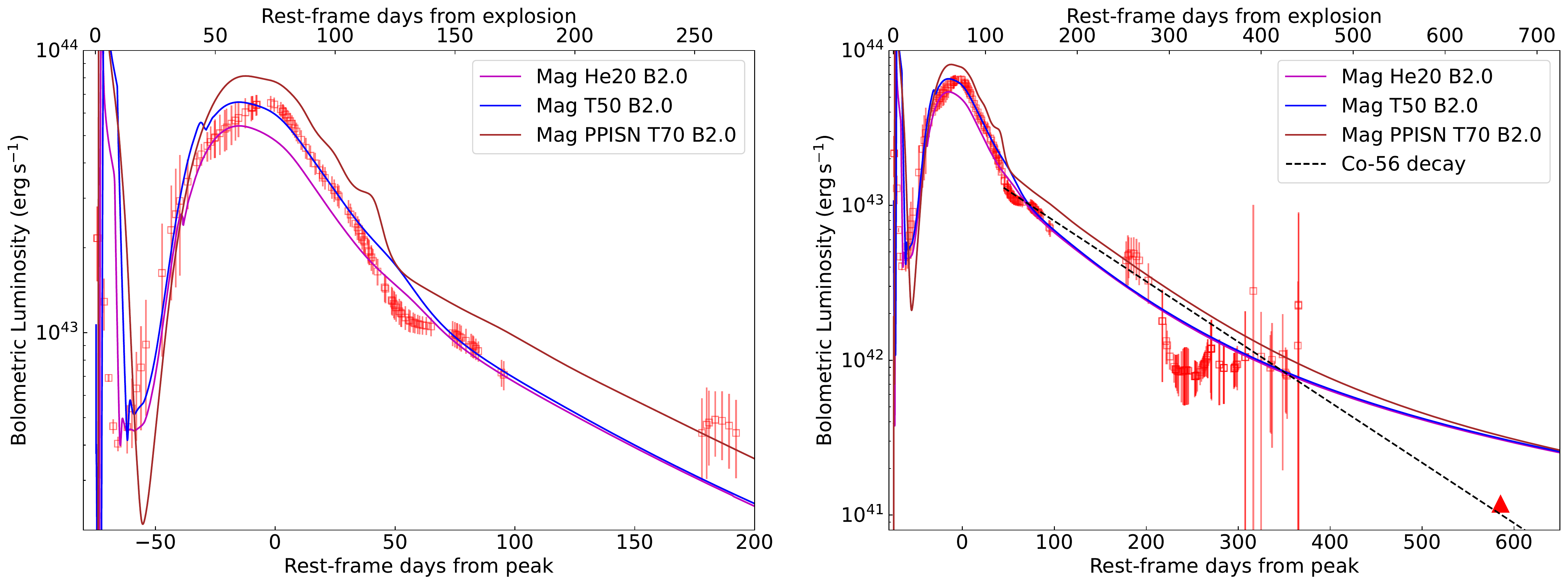}
    \caption{
    Bolometric light curve of SN\,2020wnt compared with magnetar models discussed in \textsection\ref{sec:model_comparison}.
    The red triangle denotes the lower limit of the bolometric luminosity obtained from the \textit{HST} observations of SN\,2020wnt 585 d post peak. 
    Models He20, T50, and T70 are shown; they are explosions of a massive progenitor star with energy boost from a central engine. 
    A magnetar is invoked in the three models, but an accreting black hole would serve the same purpose. 
    The magnetic field of the magnetar is $2 \times 10^{14} \rm \ G$ for all models.
    Even though the light curves are powered at peak chiefly by the magnetar, \cobalt\ decay contributes about 1/4 to 1/2 of the luminosity on the tail, hence the difference between Model T70 and Models T50 and He20.
    See detailed discussions of these models in \textsection\ref{sec:model_comparison}.
    The Left and Right panels show the light curve around peak and the entire light curve, respectively.
    These models provide markedly better fits to the data compared to the PISN models (Fig.~\ref{fig:bolometric_LC}, right). 
    }
    \label{fig:LC_magnetar}
\end{figure*}

At 585 d post peak, we derive a lower limit of the bolometric luminosity by running \texttt{SuperBol} on the 2-band \textit{HST} photometry. 
This is a lower limit because by this epoch, the near-IR contribution to the emission is expected to be significantly larger than what is captured in the optical; and that the SED in the near-IR cannot be fit by extrapolating from the optical measurements.
The SED also likely deviates significantly from a black body at this epoch.
The SN is now too faint to observe in the near-IR with a ground-based telescope; space-based observations, especially with \textit{JWST}, are required to measure the SED and the bolometric luminosity at this epoch. 
Thus, we mark this epoch with a triangle in Figs.~\ref{fig:bolometric_LC} (right) and \ref{fig:LC_magnetar}.

\subsection{Light Curve and Color Evolution}\label{sec:LC_evolution}
We identify three phases of evolution of SN\,2020wnt's light curve: early shock-cooling phase, symmetric peak, and bumpy tail. 
Immediately after the explosion, both the ZTF and ATLAS data show a quickly fading emission with $\sim$5 day decline.
Only a few epochs of data capture this feature.
This is likely the cooling emission from the close-in CSM heated by the SN shock; it has been observed in a number of SLSNe~I and SNe~Ibc \citep[e.g.,][]{leloudas2012,piro2015, nicholl2015, smith2016, taddia2016, anderson2018, angus2019, gagliano2022}.
At this phase, the emission is blue with $g-r = -0.4 \pm 0.2$~mag (Fig.~\ref{fig:photometry}, bottom).

After the initial shock-cooling emission, SN\,2020wnt rises for 75 days, peaking first in the bluer bands. 
The color remains roughly constant in this phase with $g-r \sim -0.05$~mag. 
The SN reddens to $g-r = 0.8$ around 50 days post peak. 
The spectroscopic sequence, discussed in \textsection\ref{sec:spec_evo}, will reveal that this color evolution is likely due to Fe-line blanketing. 

After 50 days post peak, the light curve appears to settle on a linear magnitude decline.
Figures~\ref{fig:photometry} and \ref{fig:bolometric_LC} show that this initial slope is consistent with what is expected from a light curve powered by the radioactive decay of \cobalt. 
Starting at 100 days post peak, there is a 70-day gap in the coverage due to the solar conjunction. 
Afterward, the SN reemerges at magnitudes roughly expected from the radioactive decay model. 

However, starting at $\sim$210 day post peak, the SN suddenly fades in all bands.
This dip occurs between 210--280 d post peak, with the minimum at $\sim$240 days where the SN fades by about 1~mag. 
Figure~\ref{fig:photometry}, bottom, shows that the SN also significantly reddens inside the dip, from the average $ (g-r)_{\rm pre-dip} = 0.31 \pm 0.05$~mag before the dip (170--210 d) to $ (g-r)_{\rm dip} = 0.69 \pm 0.05$~mag inside the dip (210--280 d).
Note that the color scatter inside and after the dip is significant due to the faintness of the SN. 
The reported values here are weighted averages, and the error bars are standard errors of the mean.
We discuss in \textsection\ref{sec:dust_CO} that this sudden dip and reddening of the light curve is likely due to dust formation and subsequent destruction.

\subsection{Explosion Properties of SN 2020wnt Assuming Radioactivity}\label{sec:exp_params}
We first estimate the ejecta mass and \nickel\ mass of SN\,2020wnt, assuming that the SN is powered solely by radioactivity. 
In the context of SLSNe~I, this calculation typically shows that the \nickel\ mass needed to explain the luminosity is too large for the ejecta mass; thus, requiring an additional power source.

We are using the analytic model of \cite{khatami2019}, which is an update of the classic \cite{arnett1982} model without the self-similarity assumption.
This model produces results that are more in agreement with numerical simulations.
In order to compute the \nickel\ mass, $M_{\rm Ni}$, produced in this SN, we use the peak luminosity of $L_{\rm peak} = (6.8 \pm 0.3) \times 10^{43} \rm \, erg\, s^{-1}$ derived previously.
The peak epoch is $t_{\rm peak} = 69 \pm 2$ days post-explosion in the rest frame. 
Since the rise time is much longer than the decay timescale of \nickel, we also consider the energy injection from the decay of \cobalt. 
The derivation in this scenario is in Appendix A1 in \cite{khatami2019}.
We can rearrange their equation (41):
\begin{align}
    M_{\rm Ni} = L_{\rm peak} \frac{\beta^2 \tau_{\rm Ni}^2}{2 \epsilon_{Ni}} \left[\right. &0.83(1-\beta \tau_{\rm Ni}) e^{-\beta \tau_{\rm Ni}} \\ \nonumber
    &+26.56(1-(1+\beta \tau_{\rm Co}) e^{-\beta \tau_{\rm Co}}) \left.\right]^{-1}
\end{align}
where $\beta$ is a parameter describing the spatial distribution of the heating function and $\tau_{\rm Ni, Co} = t_{\rm peak}/t_{\rm Ni, Co}$.
For \nickel\ and \cobalt, the decay timescales are $t_{\rm Ni} =8.8$~d and $t_{\rm Co} = 111.3$~d; and the heating rates are $\epsilon_{\rm Ni} =  3.9\times 10^{10} \, \rm erg \, s^{-1} \,  g^{-1}$ and $\epsilon_{\rm Co} =  6.8\times 10^9  \, \rm erg \, s^{-1} \, g^{-1}$, respectively.
We obtain $M_{\rm Ni} = 7.3 \pm 0.2 \ M_{\odot}$ for $\beta = 4/3$ (central heating source with constant opacity) and $M_{\rm Ni} = 5.6 \pm 0.2 \  M_{\odot}$ for $\beta = 0.9$, appropriate for SNe~Ic \citep{afsariardchi2021}.

Next, we calculate the diffusion timescale using \citet[][their equation 23]{khatami2019}.
Here we use the timescale of the \cobalt\ decay as the heating timescale, because this is the relevant decay for most of the rise: $t_s =  111.3$ d.
Solving the equation, we obtain the diffusion timescale of $t_d = 114\substack{+4 \\ -5}$~d. 
This quantity is related to the ejecta mass by rearranging \citet[][their equation 12]{khatami2019}:
\begin{equation}
     M_{\rm ej} = \frac{t_d^2 v_{\rm ej}c}{\kappa }
\end{equation}
where $\kappa$ is the opacity, $v_{\rm ej}$ is the ejecta velocity and $c$ is the speed of light. 
For the ejecta velocity, we measure the absorption trough of the \ion{Fe}{2} lines at 4924, 5018, and 5169~\AA\ near peak light (\textsection\ref{sec:spec_evo}), as these lines have been shown to well probe the ejecta velocity in SNe~Ic \citep{liu2016, modjaz2016} and have been used in SLSNe~I \citep{chen2022b}. 
The velocity is $v_{\rm ej} = 10{,}300 \ \rm km \, s^{-1}$ (see \textsection\ref{sec:opt_spec}). 
The electron scattering opacity in H-free ejecta at $11{,}000 \rm \ K$ is $\sim 0.03 \rm \ cm^2\, g^{-1}$ (Khatami 2021, private communication). 
The ejecta mass calculated is $M_{\rm ej} = 52\pm 2 \ M_{\odot} (\kappa / 0.03 \, \mathrm{cm^{2}\, g^{-1}})$.
Thus, the amount of \nickel\ required to power this SN is about 12\% of the total ejecta mass, allowing the radioactive decay of \nickel\ and its daughter species to be the sole power source of this SN.
We discuss in a later section that such a large amount of \nickel\ is very difficult to synthesize unless in a pair-instability SN, and that ultimately an additional power source is therefore still required to explain the luminosity of SN\,2020wnt.

\section{Spectroscopic Evolution}\label{sec:spec_evo}
\subsection{Optical Spectra}\label{sec:opt_spec}
Figure~\ref{fig:opt_spec} (left) shows the optical spectra of SN\,2020wnt from $-47$ days to $+359$ days from peak light.
The spectra lack hydrogen and helium at all epochs, and show features typical of SNe~Ic.
The most notable features at early times are the blueshifted absorption from \ion{C}{2} at 6580 \AA\ and 7231 \AA. 
The 6580 \AA\ absorption is comparable in strength to the \ion{Si}{2} $\lambda$6350 before peak, and both \ion{C}{2} features weaken as the SN evolves and disappear completely by about a month post peak. 
The velocity of the absorption minimum for both \ion{C}{2} lines is $-8${,}$000$ $\rm km \, s^{-1}$ at 47 d before peak, and gets slower to about $-6$,$000$ $\rm km \, s^{-1}$ by 20 d post peak. 
The \ion{Si}{2} $\lambda$6350 absorption also weakens and slows as the SN evolves, disappearing later at around 2 months post peak.
In addition, strong features in early-time spectra include the \ion{Ca}{2} H \& K lines; \ion{Na}{1} D doublet (from the SN, as well as the interstellar absorption from the Milky Way); \ion{O}{1} 7774~\AA; and \ion{Ca}{2} triplets.
There is a persistent peak at 4571~\AA, corresponding to the semi-forbidden \ion{Mg}{1}], but at early time this feature could also be due to iron line blanketing on both sides. 
There are also persistent \ion{Fe}{2} lines at 4924, 5018, and 5169~\AA, which are shown by \cite{liu2016} and \cite{modjaz2016} to probe the ejecta velocity. 
Their velocity is consistently at $-10${,}$300 \rm \ km \, s^{-1}$ until they disappear after peak.
We adopt this as the ejecta velocity of SN\,2020wnt (e.g., \textsection\ref{sec:exp_params}). 

After peak luminosity, the blue part of the spectrum fades due to line blanketing from iron-group elements, reflecting the reddening observed in photometry. 
Other spectral features do not change significantly up to 93 days post peak when the SN sets behind the Sun.
The first spectra obtained after it reemerges $\sim$100 days later show that the absorption features are largely gone. %
Forbidden lines of [\ion{O}{1}] 6300 and 6363~\AA\ and [\ion{Ca}{2}] 7292 and 7324~\AA\ appear, indicating that the density in the ejecta has sufficiently decreased for these transitions to occur. 
Permitted lines from Na, O, and Ca previously seen in absorption are now in emission.
We do not have spectra of the SN in the minimum of the light curve dip between 210--270 days (Fig.~\ref{fig:bolometric_LC}), but we do not find significant spectral changes in the optical before and after the dip.
We further discuss the spectroscopic evolution of SN\,2020wnt in comparison with other SL and SESNe in a later Subsection. 

\subsection{Near-Infrared Spectra}
\begin{figure}
    \centering
    \includegraphics[width=\linewidth]{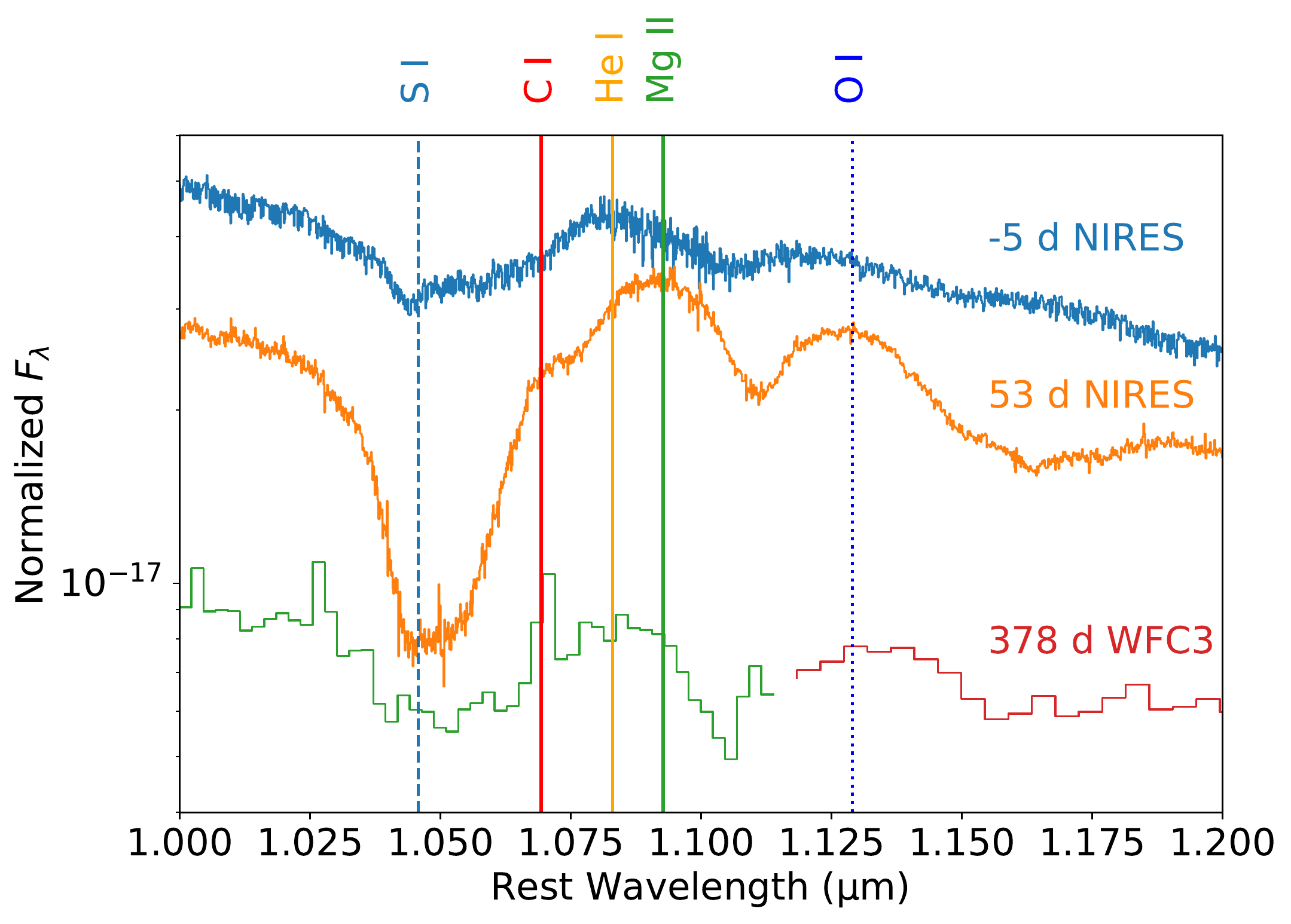}
    \caption{The 1-$\mu$m complex region of select high S/N near-IR spectra of SN\,2020wnt. Lines that potentially contribute to this complex (according to \protect\citealp{shahbandeh2021}) are marked, specifically \ion{S}{1} 1.0457 $\mu$m, \ion{C}{1} 1.0693 $\mu$m, \ion{He}{1} 1.0830 $\mu$m, and \ion{Mg}{2} 1.0927 $\mu$m. The \ion{O}{1} 1.1290 $\mu$m line is also marked. }
    \label{fig:1micron_complex}
\end{figure}

Figure~\ref{fig:opt_spec} (right) shows the near-IR spectra of SN\,2020wnt from 5 days before to 378 days after peak light.
The single near-IR spectrum pre peak ($-5$ d) is dominated by the thermal continuum with minimal spectral features. 
The line complex around 1.08 $\mu$m is present; this feature is potentially a blend between many lines, including \ion{He}{1} 1.0830 $\mu$m, \ion{C}{1} 1.0693 $\mu$m, \ion{Mg}{2} 1.0927 $\mu$m, and \ion{S}{1} 1.0457 $\mu$m \citep[e.g.,][]{shahbandeh2021}. 
Figure~\ref{fig:1micron_complex} shows the evolution of the 1-$\mu$m complex.
We do not detect the weaker \ion{He}{1} 2.0581~$\mu$m line at any epoch. 
Other spectral features present at this epoch are the \ion{O}{1} 1.1290 $\mu$m and the \ion{C}{1} 2.1259 $\mu$m lines. 

Between 52--81 days post peak, more spectral features emerge in the near-IR. 
Spectra with optical overlap (SpeX and GNIRS) show both the Ca triplet and the \ion{O}{1} 9263~\AA\ lines.
The 1-$\mu$m line complex persists as well as the \ion{O}{1} 1.1290~$\mu$m line. 
A number of peaks emerge throughout the near-IR wavelength range, resulting from multiple intermediate-mass and iron-group elements.
We identify the peak around 1.19~$\mu$m from \ion{Fe}{1}~1.1883, 1.1973~$\mu$m; the 1.257 and 1.3165~$\mu$m lines from \ion{O}{1} (potentially with some contribution from the \ion{O}{1} 1.2464 $\mu$m line); the 1.5~$\mu$m peak from \ion{Mg}{1} 1.4878 and 1.5033~$\mu$m; and the 2.135 $\mu$m feature to be a blend between \ion{Co}{2} 2.1347~$\mu$m and \ion{Mg}{2} 2.1369~$\mu$m. 
Other features in the \textit{H} band (e.g., at 1.59 and 1.68~$\mu$m) are likely due to iron-group elements.

After the solar conjunction, there are three spectra of SN\,2020wnt obtained between 241 to 378 days post peak. 
The first spectrum at 241 d is from the low-resolution prism mode of SpeX and the signal-to-noise (S/N) ratio is poor due to the faintness of the SN at this epoch ($K\approx 18.5$ mag). 
The second spectrum with Gemini/GNIRS has a better S/N.
The last spectrum from \textit{HST}/WFC3 IR grisms has the best S/N of the late-time near-IR spectra, but it only covers out to 1.6~$\mu$m. 
No new spectral features from atomic species emerge at this phase; however, the first overtone emission from CO is tentatively detected in the SpeX spectrum from 241 d, and is strongly detected in the GNIRS spectrum at 313~d.
The thermal continuum also flattens in the red part of the spectra, in agreement with near-IR photometry obtained at similar epochs suggesting dust formation (Fig.~\ref{fig:nir_sed}). 
We discuss these features in more detail in \textsection\ref{sec:dust_CO}.

The last spectrum we currently have from SN\,2020wnt is obtained with \textit{HST}/WFC3 near-IR grisms at 390 d post peak. 
The continuum is still clearly present at this epoch and the strong 1-$\mu$m complex still has some P-Cygni absorption.
We note, regardless, that the peak of the 1-$\mu$m complex is now at the rest velocity of \ion{He}{1}.
A later observation will allow us to more confidently identify the species responsible for the 1-$\mu$m complex. 
We further compare this spectrum to models of massive core-collapse SNe and PISNe in \textsection\ref{subsec:nir_nebular_comparison}.

\subsection{Formation of Carbon Monoxide and Dust}\label{sec:dust_CO}
SN ejecta form molecules as they expand and cool, starting with simple molecules and progressing to more complex molecules and dust grains as the temperature drops.  
Dust formation in SN ejecta is of particular interest because it has the potential to explain the large dust mass observed in the early universe \citep[e.g.,][and references therein]{gall2011}, before other dust-forming objects such as asymptotic giant branch (AGB) stars had sufficient time to evolve.
While massive stars may produce dust in their outflows \citep[e.g.,][]{lau2021}, their death as CCSNe is potentially the dominant channel \citep[e.g.,][]{gall2018}.
Various observations of SN\,1987A and SN remnants in the radio and sub-millimeter demonstrate that $\sim$0.1 $M_\odot$ of dust can eventually form in the aftermath of a CCSN \citep[e.g.,][]{rho2008, rho2009, barlow2010, indebetouw2014, matsuura2015, dwek2015, lau2015, rho2018b}.
However, measurements from near- to mid-IR photometry and spectroscopy of CCSNe from months to years post-explosion \citep[e.g.][]{szalai2013,tinyanont2016, szalai2019, tinyanont2019b, rho2021, li2022} show that the mass of newly formed at these early times remain modest around {$10^{-5}$--$10^{-3} \, M_\odot$}. 
(Strongly interacting SNe show much larger dust mass, but most of it is likely pre-existing in the CSM, \citealp{fox2011,fox2013}.)
In the coming years, \textit{JWST} near-to-mid-IR spectroscopy will paint a much more complete picture of the evolution of dust mass in local CCSNe. 
However, a question remains whether primordial CCSNe with massive progenitors and small metallicity produce molecules and dust in the same way. 

\begin{figure}
    \centering
    \includegraphics[width=\linewidth]{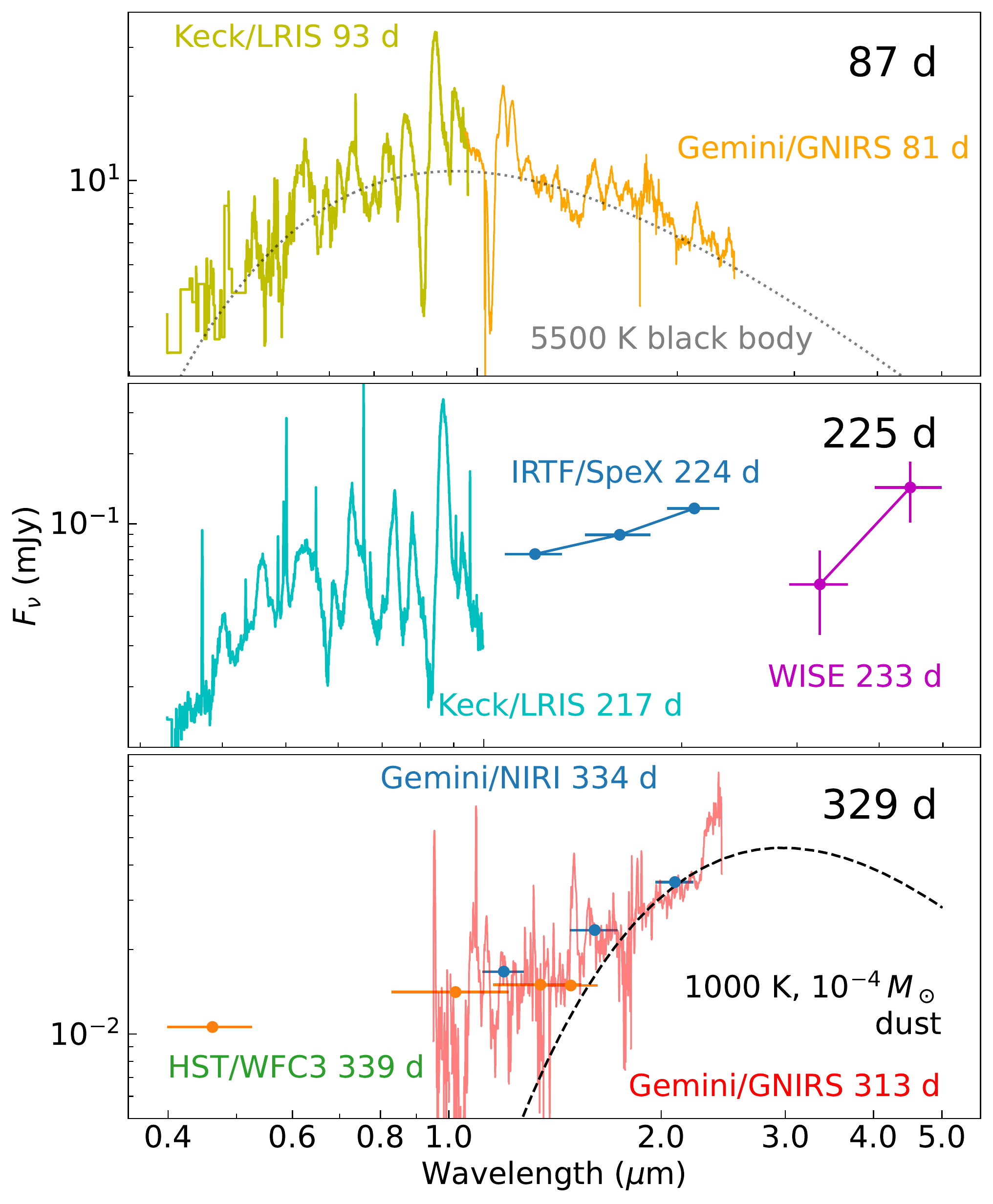}
    \caption{Evolution of the SED of SN\,2020wnt. Multi-band photometry and spectroscopy from three epochs, 87, 225, and 329~d post peak, are shown in the top, middle, and bottom panels, respectively. 
    The data are corrected for the Galactic extinction. 
    At $\sim$87~d, the SED can be explained as 5500~K black body radiation.
    At 217--233~d, the near-IR contribution becomes significant, and there is a sign of CO formation in the W2 channel of \textit{WISE}. 
    The data from different epochs in this panel are scaled based on the $r$-band photometry because this is the epoch during which the light curve declines quickly. 
    We note that the fluxes from SpeX and \textit{WISE} do not match up, likely because the SN was quickly fading around this epoch.
    The near-IR contribution strengthened and by $\sim$329 d, a continuum from hot, 1000~K dust is needed to explain the SED. 
    The inferred dust mass, assuming a simplistic 0.1~$\mu$m carbonaceous grain population, is $10^{-4} \, M_\odot$. 
    The first overtone emission of CO is also clearly detected around 2.3~$\mu$m. 
    }
    \label{fig:nir_sed}
\end{figure}

Prior to dust formation, SN ejecta form more simple molecules since these less complex chemical species can withstand higher temperature and pressure.
Carbon monoxide (CO) is the first molecule to form due to its strong molecular triple bond.
In SN\,2020wnt, the first sign of CO emerges around 230 d post peak (around 300 d post explosion). 
The near-IR spectrum from 241 d shows a tentative detection of the first overtone band of CO starting at 2.3 $\mu$m, a feature which strengthens and is clearly detected at 313 d (Fig.~\ref{fig:nir_sed}). 
Further, the \textit{WISE} photometry from 233 d shows a strong excess in the W2 (4.6 $\mu$m) channel, which is centered on the CO fundamental band. 

CO emission has been observed in CCSNe of all types with late-time near-IR spectra \citep[e.g.][]{spyromilio1988, gerardy2002, rho2018,tinyanont2019b, davis2019, rho2021, shahbandeh2021}.
The formation happens in the C/O rich layer (He-burning product) of the ejecta \citep{sarangi2015, jerkstrand2017}, equally depleting carbon and oxygen from the gas, leaving the remaining elements for further dust formation.
C-rich materials can form different forms of carbonaceous dust such as amorphous carbon and graphite.
O-rich materials further form oxides such as $\rm SiO$ and $\rm SiO_2$, and eventually silicate dust. 
Other O-rich molecules, e.g., CO$_2$ and O$_2$ may eventually form as well.
CO emission also cools the ejecta to temperatures appropriate for dust formation ($\lesssim$1500 K) \citep[][and references therein]{sarangi2018}.
Detecting this feature in the majority of SLSNe is impossible from the ground because the 2.3 $\mu$m band head gets redshifted out of the ground-based near-IR band ($\sim$2.5~$\mu$m) at a redshift of only $z = 0.087$.
The CO detection in SN\,2020wnt presents a strong case for near- to mid-IR observations of SLSNe with \textit{JWST} to start to map the chemical evolution in the ejecta of these explosions.

In addition to the CO emission, we detected a rising continuum, presumably from dust grains. 
During the strong light curve dip around 220 d post peak, the spectral energy distribution (SED) with near-IR photometry from SpeX and \textit{WISE} shows an emerging dust thermal continuum (though the W2 point is likely contaminated by CO emission).
At around 330 d post peak, the SED can be explained by adding a thermal component of hot dust with $T = 1000 \, \rm K$ and $M = 10^{-4} \, M_\odot$, assuming a simple population of 0.1 $\mu$m carbonaceous grains. 
The available data are insufficient to distinguish different dust compositions and grain sizes; mid-IR spectroscopy covering the silicate features at 10 and 18.9 $\mu$m is needed.

To determine whether dust formation explains the sudden decline in the luminosity, and reddening, at around 210 d post peak, we compute the time-dependent reddening of SN\,2020wnt by fitting its SED around the dip assuming a constant temperature of 8,000~K, the temperature found by \texttt{SuperBol} for this epoch. 
Note that this temperature was derived assuming the Galactic extinction with $E(B-V) = 0.42$. 
The extra time-dependent extinction found by this analysis is assumed to be intrinsic to the SN.

During the dip, the maximum total reddening is $E(B-V) \approx 0.69$~mag, corresponding to $A_{r} =1.7$~mag (an increase from $A_{r} =1.0$~mag just from the MW).  
The decrease in $r$-band flux during the dip compared to the interpolated light curve from pre- and post-dip photometry is 1.84~mag, suggesting that dust formation can roughly account for the flux decrement at these times.
To explain the recovery out of the dip later, the dust can be destroyed by the shock front or there is a strong asymmetry in dust formation which allows the dusty part of the ejecta to obstruct a smaller solid angle at later times.

Molecule and dust production in SNe remain poorly constrained due to the lack of IR observations.
For SLSNe~I, there has only been one event with signs of dust formation detected with \textit{Spitzer} \citep[SN\,2018bsz;][]{chen2021}, and three more with \textit{WISE} \citep{sun2022}.
At least for SN\,2018bsz, CSM interactions may have played a role in illuminating pre-existing dust or forming new dust in the post-shock ejecta \citep{pursiainen2022}.
Incidentally, SN\,2018bsz also showed \ion{C}{2} in early-time spectra, but it also had other typical SLSNe~I features (\ion{O}{2} absorption and a blue continuum). 
With model predictions of continued dust formations in these SLSN remnants aided by the pulsar emission from the magnetar \citep{omand2019}, and the possibility that these superluminous explosions are more common in the early universe, further space-based IR observations are warranted.

\subsection{Optical Spectral Similarities to Stripped-Envelope Supernovae Near Peak}
To quantitatively show that SN\,2020wnt is spectroscopically more similar to SESNe (and to an extent SNe~Ia) near peak than to engine-driven SLSNe, we follow the methodology outlined in \citet{quimby2018} to quantitatively determine a spectrum's classification. 
We compare the optical spectra of SN\,2020wnt to a large template library of SN spectra from \citet{quimby2018}\footnote{Accessible at \url{https://github.com/rmquimby/2018ApJ...855....2Q}}
using \texttt{Superfit} \citep{howell2006}.
We similarly limit the wavelength range used to 3900--7000~\AA, to avoid biases due to the incomplete wavelength coverage of many of the templates.
We do not perform this analysis on our spectra with a gap in this wavelength region for the same reason.
We limit the redshift range searched by \texttt{Superfit} to keep the runtime reasonable.
All other setups are the default.
Given an observed spectrum, \texttt{Superfit} provides a list of best-match templates ranked inversely by the $\chi^2$ value of the fit. 
We then compute a classification score for each class by averaging the indices of the top five matches for each class. 
For instance, if the first five SLSN~I templates that match a given spectrum have indices 0, 1, 3, 5, and 6, the SLSN~I classification score for this spectrum is 3.
The class with the smallest score is thus the best match.

\citet{quimby2018} showed that the difference between classification scores of two classes (e.g., SN Ic - SLSN I) can reliably distinguish SLSNe from other types of SNe (and also different SNe subtypes from each other).
This result establishes that SLSNe form a distinct spectroscopic class from other SESNe, regardless of the luminosity, and that some SLSNe peak at lower luminosities than the traditional cutoff at $-21$~mag.

\begin{figure*}
    \centering
    \includegraphics[width = 0.7\linewidth]{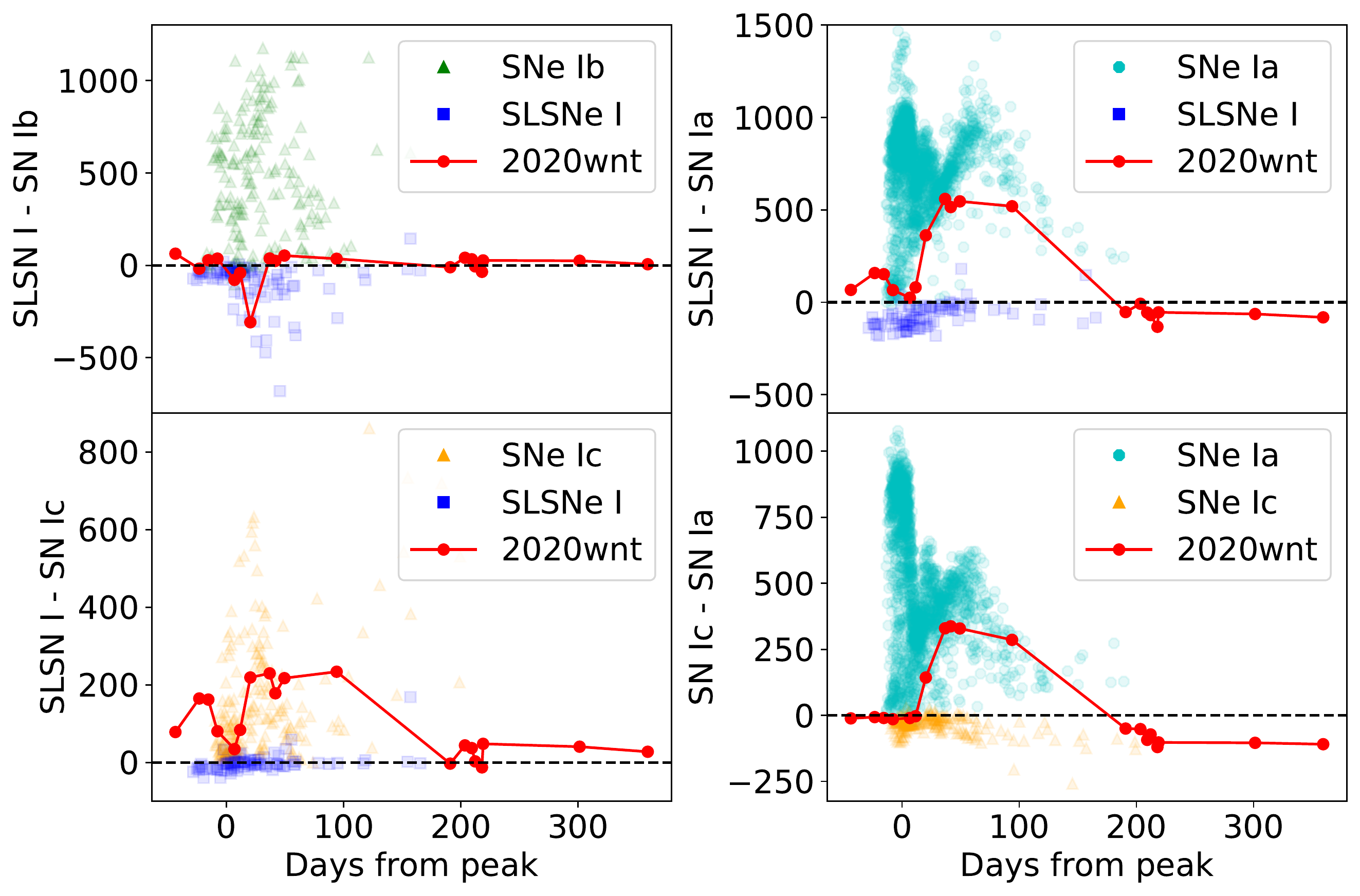}
    \caption{Classification score differences of SN\,2020wnt, in comparison with other SNe and SLSNe in the literature. 
    The literature data come from \protect\cite{quimby2018}, and these plots resemble their Figures 3 and 4.
    Clockwise starting from top left, we plot the SLSN~I $-$ Ib, SLSN~I $-$ Ia, Ic $-$ Ia, and SLSN~I $-$ Ic scores.
    These scores allow us to quantitatively determine if a given spectrum more resemble one class or another in a pairwise fashion.
    For instance, in the SLSN~I $-$ Ib plot, most known SNe Ib have positive score difference while SLSNe~I are negative.
    SN\,2020wnt's spectra lie close to zero, meaning that both classes are unlikely. 
    From these plots, we can conclude that the spectra of SN\,2020wnt resemble those of SNe Ic at all phases, and also SNe Ia at around 1 month post peak. 
    We provide further discussion in the text. 
    }
    \label{fig:type_score}
\end{figure*}

Figure~\ref{fig:type_score} shows the SLSN~I$-$SN Ia, Ib, Ic, and SN Ia$-$Ic scores of SN\,2020wnt, plotted on top of other SNe of known subtype from \cite{quimby2018}. 
The left column shows the SLSN~I$-$SN Ib and Ic scores, while the right column shows SLSN~I$-$SN Ia and SN Ic$-$Ia scores.
These plots demonstrate that the score differences reliably distinguish SNe with different known spectral types. 
SN\,2020wnt lies roughly on the zero line for the SLSN~I$-$SN Ib score, showing that it unlikely belongs to either class.
The SLSN~I$-$SN Ia and Ic plots show that SN\,2020wnt better resembles either SNe~Ia or Ic than it does SLSNe~I. 
Lastly, in the SN Ic$-$Ia plot, SN\,2020wnt better resembles SNe Ic at most epochs, except for the epochs between 20--93~d post peak. 
These plots also show that spectra with most distinguishing power (largest score differences) are those from between 30 and 100~d post peak, around the time when SN\,2020wnt enters the nebular phase, revealing its inner ejecta. 
After around this phase, the analysis becomes difficult due to the lack of template spectra at these epochs.

To visualize this similarity, Fig.~\ref{fig:spec_comparison} shows select high S/N spectra of SN\,2020wnt pre peak (-43 d), near peak (+6 d), early nebular phase (49 d), and later in the nebular phase (+359 d) in comparison with SN\,2007gr (Ic, \citealp{valenti2008, hunter2009}), SN\,2015bn (SLSN I, \citealp{nicholl2016a, nicholl2016b}), and SN\,2017gci (SLSN I, \citealp{fiore2021}) at comparable epochs.
We include the mean nebular SLSN and SN~Ic spectra from \cite{nicholl2019} at late time, in order to discuss SN\,2020wnt in context of other SLSNe and SNe~Ic. %
We also include a spectrum of SN~Ia (91T-like) 1999dq \citep{blondin2012} at 29 d post-peak, which is the best match to the early nebular spectra of SN\,2020wnt.

The pre-peak spectra (top two panels in Fig.~\ref{fig:spec_comparison}) clearly show the lack of both a hot continuum and a series of \ion{O}{2} absorption, which are associated with SLSNe~I. 
The ejecta of SLSNe~I get continuously heated by the central magnetar, allowing them to remain hot and producing these spectral features all the way from explosion to even after peak light \citep[][and references therein]{nicholl2021}. 
The absence of this temperature excess in SN\,2020wnt requires that a central engine either does not exist or that its effects are concealed until well after peak.

At around 50 d post peak when the light curve of SN\,2020wnt falls onto the approximately linear decline, the spectra are best matched not by SNe~Ic, but SNe~Ia. 
The middle panels of Fig.~\ref{fig:spec_comparison} and Fig.~\ref{fig:type_score} show that while the spectrum of SN\,2020wnt remains distinct from those of SLSNe~I, it now more resembles that of a SN~Ia, rather than a SN~Ic at a comparable phase.
In fact, during this phase, the first SN~Ic match we obtain from \texttt{Superfit} is number 320 on the list; everything above is a SN~Ia spectrum.
This result suggests that the composition of the outer ejecta of SN\,2020wnt and the temperature (thus atomic excitation states) are similar to those of SNe~Ia at this epoch.
Detailed modeling of this phase of spectra is beyond the scope of this paper.

\subsection{Late Nebular Spectra Reveal a Magnetar}

\begin{figure}[ht]
    \centering
    \includegraphics[width=0.85\linewidth]{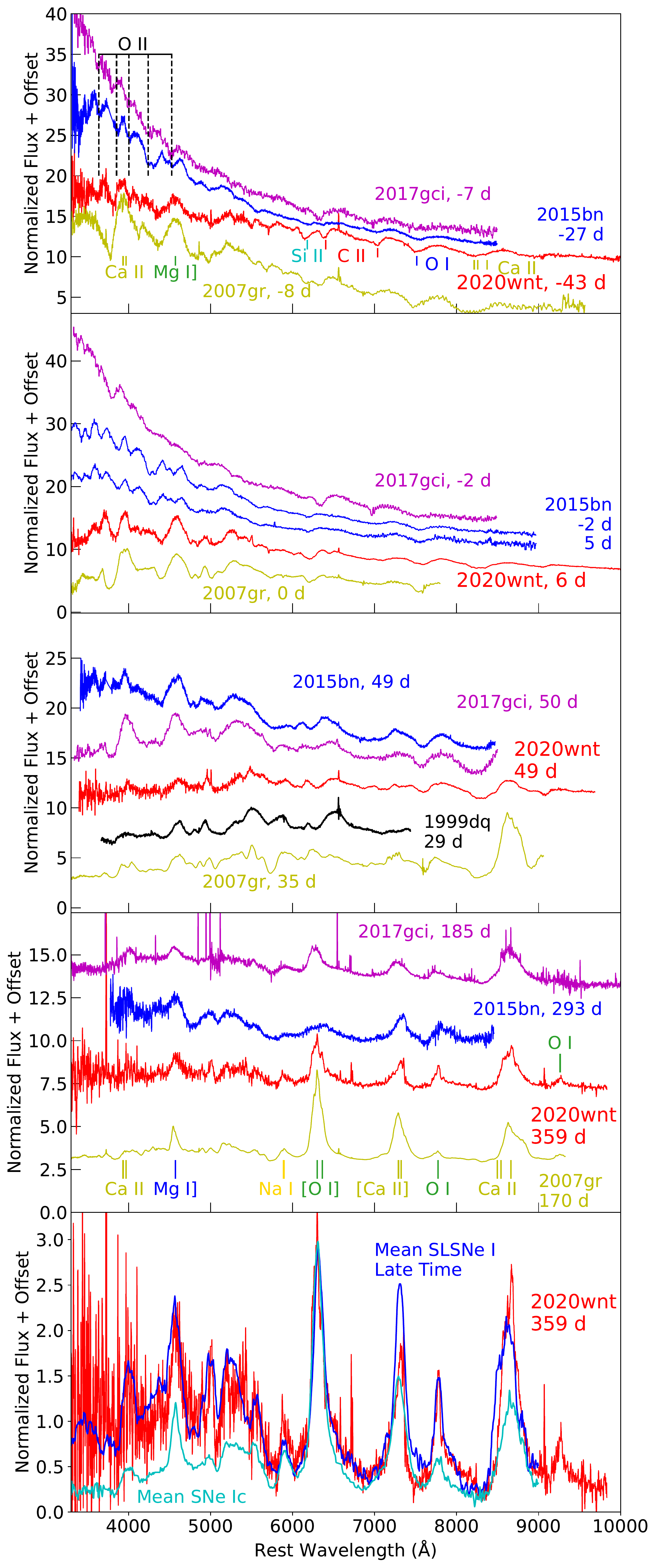}
    \caption{\textbf{Top four panels:} Optical spectra of SN\,2020wnt at $-43$, 6, 49, and 359 days from peak, compared with those of SN~Ic 2007gr \protect\citep{hunter2009}, and SLSNe 2015bn \protect\citep{nicholl2016a, nicholl2016b} and 2017gci \protect\citep{fiore2021} at comparable epochs. For the 49 d epoch, the best-matched spectrum, SN\,1999dq (Ia 91T-like) \protect\citep{blondin2012} from 29 d post peak, is also shown.
    \textbf{Bottom:} The last optical spectrum of SN\,2020wnt at 359 d compared with the mean late-time SLSN~I and SN~Ic spectra from \protect\cite{nicholl2019}. 
    The two mean spectra plotted are similar to their Figure 8.
    }
    \label{fig:spec_comparison}
\end{figure}

\cite{nicholl2019} compared nebular spectra of SLSNe~I and SNe Ic and demonstrated that while they are less distinguishable than spectra near peak, there are a number of subtle differences between nebular spectra of SLSNe~I and SNe~Ic, potentially due to magnetar heating in the former class. 
First, the mean nebular spectrum of SLSNe~I has more luminous recombination \ion{O}{1}~7774~\AA\ emission with a relatively narrow profile, indicating that it originates in the inner ejecta. 
This line demonstrates that the inner ejecta of SLSNe~I remain ionized at late times due to the power-law nature of magnetar heating luminosity (as opposed to exponential decay for radioactivity). 
Second, the mean nebular spectrum of SLSNe~I shows an emission feature at 5000~\AA, which \cite{nicholl2019} identified as [\ion{Fe}{2}] and [\ion{O}{3}]; a magnetar is again required to keep the inner ejecta at such a high ionization state at late times. 
Third, SLSNe~I show elevated flux in the blue part of the spectrum, due to a large iron-group element synthesized in the magnetar-driven explosion. 
This blue pseudo-continuum is similar to what is observed in SNe~Ic-BL, like SN\,1998bw, which are known to produce a relatively large amount of iron-group elements (0.3--0.7 $M_\odot$ in the case of SN\,1998bw; \citealp{galama1998, sollerman2002}). 
The bottom panel of Fig.~\ref{fig:spec_comparison} compares the last optical spectrum of SN\,2020wnt with the mean late-time spectra of SLSNe~I and SNe~Ic, both from \cite{nicholl2019}.
The late nebular spectrum of SN\,2020wnt shows all these three features, and much more closely resembles the mean SLSN~I spectrum.

From these spectra, we conclude that SN\,2020wnt harbors a magnetar.
However, unlike in other SLSNe, the magnetar is hidden inside the optically thick ejecta near peak, and does not imprint perceptible spectroscopic signatures. 
This finding is quite similar to the case of iPTF15dtg, whose peak spectra clearly resemble those of normal SNe Ic \citep{taddia2016} but late-time observations show signs of magnetar power \citep{taddia2019} in both the light curve evolution and the optical spectra. 
The implication here is that we may not be able to use peak-time spectra alone to select magnetar-driven explosions from a population of H-poor CCSNe.
In some cases, like SN\,2020wnt, spectroscopic signatures of the magnetar are not apparent until $\sim$a year post peak. 
A survey of nebular spectra of SESNe is needed to determine the fraction of these explosions significantly powered by a central engine.

\subsection{Near-Infrared Spectral Comparison}
\begin{figure}
    \centering
    \includegraphics[width=\linewidth]{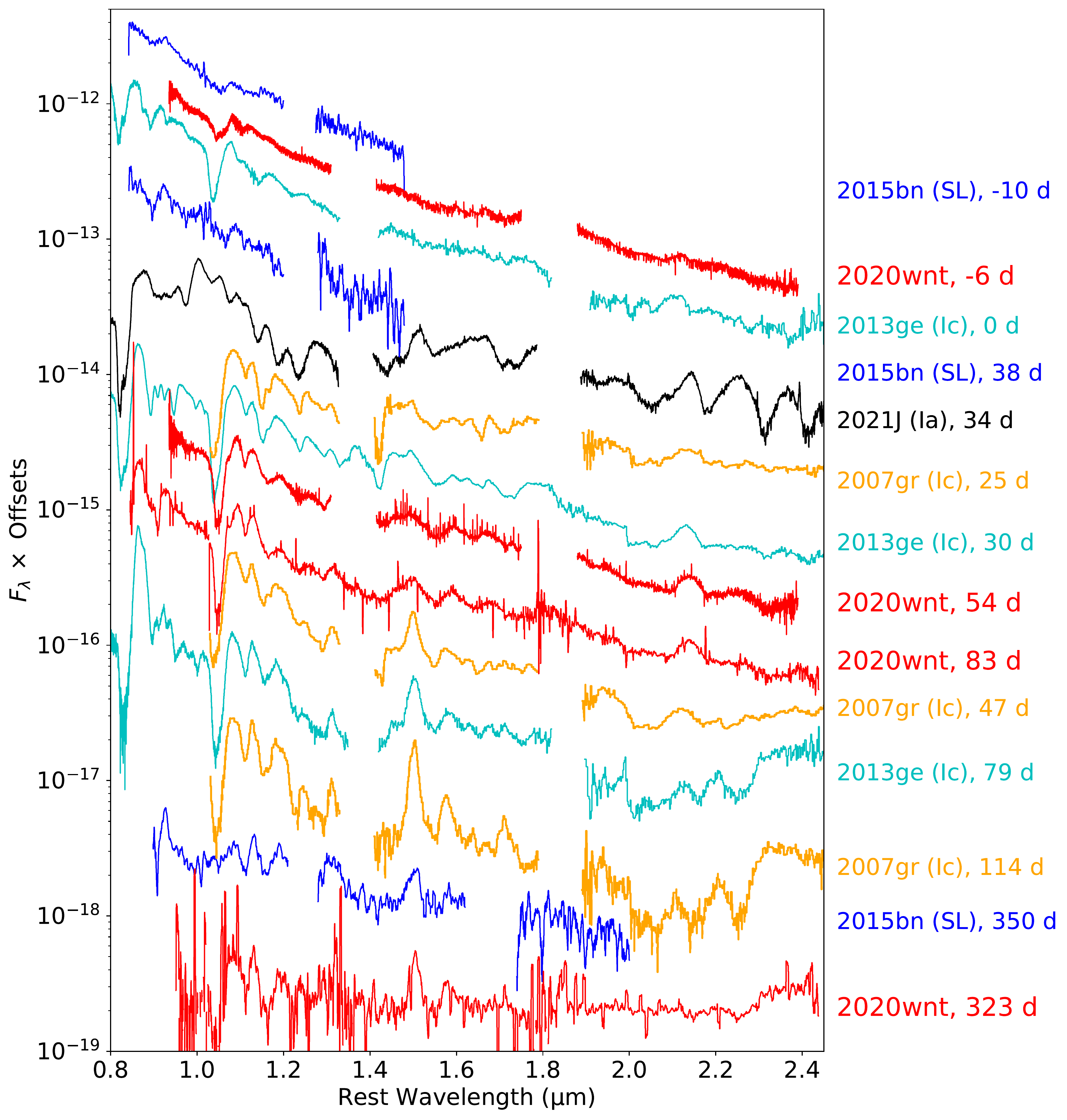} 
    \caption{Comparison between near-IR spectra of SN\,2020wnt at $-6$, 54, 83, and 323 d from peak and spectra of SNe Ic 2007gr \protect\citep{hunter2009} and 2013ge \protect\citep{shahbandeh2021}; SLSN\,2015bn \protect\citep{nicholl2016a, jerkstrand2017}; and SN Ia 2021J from our IRTF observations. 
    Spectra of other SNe are provided at comparable epochs provided availability. }
    \label{fig:IR_spec_comparison}
\end{figure}

Fig~\ref{fig:IR_spec_comparison} compares the near-IR spectra of SN\,2020wnt at $-6$, 54, 83, and 323 d from peak with the spectra of SNe Ic 2007gr \citep{hunter2009} and 2013ge \citep{shahbandeh2021}; SLSN 2015bn \citep{nicholl2016a}; and SN~Ia 2021J about 1 month post peak (this work).
These are comparable in epoch to the optical comparison in Fig.~\ref{fig:spec_comparison}. 
In the near-IR, SN\,2020wnt spectra also better resemble those of SNe~Ic compared with those of SLSNe~I. 
The comparison is more difficult because there are only a handful of SNe Ic with multi-epoch near-IR spectra and there are only a few near-IR spectra of any SLSNe~I (e.g., Gaia16apd, \citealp{yan2017} and LSQ14an, \citealp{jerkstrand2017}).
Note that the spectra of SN\,2015bn have smaller wavelength coverage due to the significant redshift and the lack of \textit{K}-band observations in the first two epochs. 
Its ground-based spectra never cover the CO bandhead, despite the relatively modest redshift for a SLSN, highlighting the need for \textit{JWST} to discern molecule (and dust) formation in SLSNe. 

The most notable difference between near-IR spectra of SLSNe and those of SNe~Ic and SN\,2020wnt is the 1-$\mu$m complex. 
The spectra of SN\,2015bn and Gaia16apd lack the deep P-Cygni absorption from this spectral feature at all epochs. 
This difference is most stark when comparing the SN\,2015bn spectrum from 350 d post peak and that of SN\,2020wnt 323 d post peak. 
While the peak at 1.08 $\mu$m (likely due to \ion{He}{1} at these epochs), the \ion{O}{1}~1.1290~$\mu$m line, and the \ion{Mg}{1}~1.5033~$\mu$m line are present in both spectra, the deep P-Cygni absorption associated with the 1.08 $\mu$m feature, observed in SN\,2020wnt at all epochs, is missing from the spectra of SN\,2015bn. 
The presence of the P-Cygni feature here also suggests that the ejecta of SN\,2020wnt are more optically thick than those of SLSNe, even at this late phase. 

Lastly, while the early nebular spectra of SN\,2020wnt resemble those of SNe~Ia from about 20--30 d post peak, the near-IR spectra are completely different. 
We show an IRTF spectrum of SN\,2021J at 34 d post peak, which has completely different spectral features compared to SN\,2020wnt and SNe~Ic. 
(This spectrum was obtained with IRTF/SpeX with the same observation and data reduction details as those for the SN\,2020wnt's SpeX spectra.) 
Most features are due to Fe-group elements, which are abundant and are the main coolants in SNe~Ia. 
The lack of these features in SN\,2020wnt does not necessarily imply a small amount of Fe-group elements.
(Indeed, PISN nebular spectral models, e.g., \citealp{jerkstrand2016, mazzali2019} all show modest Fe-group emissions in the near-IR despite the large amount of Fe-group elements synthesized in those explosions.)
With its massive ejecta, the Fe/Ni-rich part may still be embedded at these early nebular epochs. 
Even at later times when the inner ejecta are visible, the primary coolant could still be other atomic and molecular lines. 
Further detailed modeling of the IR spectra is needed to constrain the amount of Fe-group elements from the ejecta.

\section{Radio Limit on Circumstellar Medium Density}\label{sec:csm_limit}
Radio observations of SN\,2020wnt allow us to place a limit on the wind density around the progenitor star. 
Note that this limit only applies to the wind density around the shock location at the epoch of observations; it does not preclude a dense CSM close to the SN. 
Because of the lack of narrow lines from the ionized CSM in any phase, we assume that the radio absorption is dominated by synchrotron self-absorption (SSA), and not free-free absorption in the CSM. 
We use Equation (12) in \cite{chandra2018} (Equation 14 in \citealt{chevalier1998}) to convert the observed flux limits to upper limits on the magnetic field in the shock front of SN\,2020wnt. 
\begin{equation}
    B = 0.58 a^{\frac{-4}{19}} \left(\frac{f}{0.5}\right)^{\frac{-4}{19}} \left(\frac{F_{\rm peak}}{\mathrm{Jy}}\right)^{\frac{-2}{19}} \left(\frac{d}{\mathrm{Mpc}}\right)^{\frac{-4}{19}} \left(\frac{\nu_{\rm peak}}{5 \, \mathrm{GHz}}\right) \, \rm G
\end{equation}
where $a = \epsilon_e/\epsilon_B$ is the equipartition parameter (ratio between the shock energy deposited in relativistic electrons and magnetic field), $f$ is the volume filling factor of the CSM, and $F_{\rm peak}$ and $\nu_{\rm peak}$ are the flux and the frequency of the peak in the radio SED. 
This formula assumes the power index of the electron energy distribution, $N(E) \propto E^{-p}$, to be $p = 3$.
Equating the magnetic pressure to the shock ram pressure, $B^2/8\pi = \zeta  \rho_{\rm wind} v_{\rm shock}^2$, where $\zeta < 1$ is a numerical factor, one can derive Equation (13) in \cite{chandra2018}:
\begin{align}
    \dot{M} = \frac{6 \times 10^{-7} \  M_\odot \, \mathrm{yr^{-1}} }{m^2 \zeta} 
        & \left( \frac{B}{1 \ \mathrm{G}} \right)^2  \\ \nonumber & \left(\frac{t}{100\ \mathrm{d}}\right)^2 \left(\frac{v_{\rm wind}}{10 \ \mathrm{km\, s^{-1}}}\right)
\end{align}
where $m$ is the deceleration factor, which we assume to be 1 (constant velocity shock). 
We rearrange this in terms of $\dot{M}/v_{\rm wind}$, the wind density parameter, since we have no constraint on the wind velocity:
\begin{equation}
    \frac{\dot{M}}{v_{\rm wind}} = \frac{3.8 \times 10^{13} \ \mathrm{g \, cm^{-1}} }{m^2 \zeta} 
         \left( \frac{B}{1 \ \mathrm{G}} \right)^2   \left(\frac{t}{100\ \mathrm{d}}\right)^2 
\end{equation}

From these equations, we note that the magnetic field, and thus the wind density parameter, only depends weakly on the observed radio flux. 
What sets the limit, is the break frequency of the radio SED, $\nu_{\rm peak}$.
For SN\,2020wnt, there is no detection at 335 d post explosion, so we use the flux upper limit at the lowest observed frequency to put a constraint on the wind parameter of its CSM. 
This is $F_{\rm peak} < 5.4 \ \mu \rm Jy$ in the C band, at 5 GHz. 
Because of the also deep non-detection ($< 6.5 \ \mu \rm Jy$) in the Ku band, at 15 GHz, we assume that $\nu_{\rm peak} < 5 \ \rm GHz$.
While the peak frequency could be much higher and the emission is completely self-absorbed at our observed frequency, the CSM density would be high, which is inconsistent with the lack of narrow lines in the optical at any epoch.
The high CSM density scenario is also unlikely at the epoch of observation. 

We thus get an upper limit on the magnetic field of $B < 0.63 \ \rm G$, and the wind density parameter of $\dot{M}/v_{\rm wind} < 1.7 \times 10^{14} \ \rm g \, cm^{-1}$, or $\dot{M} < 1.3 \times 10^{-4} \ M_\odot \, \mathrm{yr^{-1}} (v_{\rm wind}/500\ \rm km\, s^{-1})$.
This upper limit rules out dense and extended CSM seen in strongly interacting SNe, which have $\dot{M}/v_{\rm wind} \gtrsim 5 \times 10^{15} \ \rm g \, cm^{-1}$ \citep{smith2017}; and is consistent with the progenitor of SN\,2020wnt being a massive stripped star.

From this wind density parameter limit, we can also calculate a limit on the total luminosity from CSM interactions: $L_{\rm CSM} \sim 1/2 \  \dot{M}/v_{\rm wind} v_{\rm shock}^3$.
With a shock velocity between $10${,}$000 \ \rm km\, s^{-1}$ (ejecta velocity from the \ion{Fe}{2}) and $20${,}$000 \ \rm km\, s^{-1}$ (velocity of the outer shock from the lowest mass explosion model in \textsection\ref{sec:model_comparison}), the interaction luminosity ranges from $9\times 10^{40}$ -- $7 \times 10^{41} \ \rm erg \, s^{-1}$.
At the brightest, the interaction luminosity does not significantly contribute to the bolometric luminosity of SN\,2020wnt until at least 500 d post-peak.

\section{Light Curve Modeling}\label{sec:model_comparison}

\begin{deluxetable*}{ccccccccccccc}
\tablecaption{Models for SN\,2020wnt}
\tablehead{ \colhead{Model}  &
            \colhead{${J_{0}}$} &
            \colhead{${M_{\rm preSN}}$} &
            \colhead{${M_{\rm Fe}}$}  &
            \colhead{${M_{\rm cut}}$}  &
            \colhead{${E_{\rm ej}}$}  &
            \colhead{${M_{\rm ej}}$}  &
            \colhead{${M_{\rm Ni}}$}  &
            \colhead{${J_{\rm Fe}}$} &
            \colhead{${E_{\rm puls}}$}  &
            \colhead{${B_{\rm puls}}$}  &
            \colhead{${f_{\rm ej,  puls}}$} &
            \colhead{Comments}  
         \\
            \colhead{}  &
            \colhead{[$10^{53}$ erg s]}  &
            \colhead{[\Msun]}  &
            \colhead{[\Msun]}  &
            \colhead{[\Msun]}  &
            \colhead{[$10^{51}$ erg]}  &
            \colhead{[\Msun]}  &
            \colhead{[\Msun]}  &
            \colhead{[$10^{49}$ erg s]}  &
            \colhead{[$10^{51}$ erg]}  &
            \colhead{[$10^{14}$ G]}  &
            \colhead{}  &
            \colhead{}  
             }
\startdata
C95   &  0  &  95.0  &  0.0  & 0.0  &  37.2  &  95.0  &  2.85 &  0.0 & 0.0 & -- & -- & CO core; no $\dot M$; PISN \\
C100   &  0  & 100.0  & 0.0   & 0.0  &  43.7  &  100.0 &  5.13 &  0.0 & 0.0 & -- &-- & CO core; no $\dot M$; PISN  \\
C105  &  0  & 105.0  & 0.0   & 0.0  &  48.6  &  105.0 &  10.6 &  0.0 & 0.0  & -- & -- & CO core; no $\dot M$; PISN  \\
He20  &  0  & 16.95  & 1.94  & 2.00 &  5.0   &  14.94 &  0.88 &  0.0 & 1.2  & 2.0 & 0.54  &He core; no rot.; CSM  \\
T50  &  1.5 & 28.36  & 2.41  & 2.50  & 15.0  &  25.86 &  0.85 &  2.4 & 2.0  & 2.0 & 0.41 & ZAMS; Rapid rot.; CSM  \\
T70  &  3.5 & 41.84  & 2.70  & 2.50  & 19.1  &  39.34 &  1.71 &  3.9 & 2.0  & 2.0 & 0.34 & ZAMS; Rapid rot.; PPISN   \\
\enddata
\tablecomments{The initial mass is the numerical part of the model  name, and $J_0$ is its total angular momentum. Models with $J_0$ = 0
  are not rotating. $M_{\rm Fe}$ is the mass of the presupernova iron
  core, and $M_{\rm cut}$, the location of the piston used to eject
  all exterior mass. Explosion energy includes work done by the piston
  and magnetar minus presupernova binding energy and energy emitted as
  light. $M_{\rm ej}$ is the total mass ejected and equals $M_{\rm
    preSN}$ - $M_{\rm cut}$. $J_{\rm Fe}$ is the final angular momentum
  inside the mass cut. $M_{\rm Ni}$ is the ejected mass of $^{56}$Ni.
  $E_{\rm puls}$ is the initial rotational energy of the embedded
  magnetar with the magnetic field of ${B_{\rm puls}} = 2 \times 10^{14} \ \rm G$, when one is present. 
  ${f_{\rm ej,  puls}}$ is the mass fraction of the ejecta directly heated by the magnetar.
  ``ZAMS'' in ``comments'' indicates that the star began its
  life as a hydrogen-rich star on the main sequence. Other initial models
  were He or CO cores. Models with a ``CSM'' were surrounded by a
  pre-SN stellar wind.}  \label{tab:new_models}
\end{deluxetable*}
SN\,2020wnt was no ordinary SN. 
Non-rotating, neutrino-powered, CCSNe generate inadequate energy and \nickel\ to explain its evolution \citep{Ert20}. 
If collisionally powered at peak, a large amount of matter would be required at distances $\sim10^{15}$ cm. The radio limits on mass loss after the peak would require discontinuous mass ejection. 
Similarly, the dip in the light curve at $\sim20$ days would be difficult to explain without some sort of explosive shell event months to years before the star died. Instead, the broad smooth light curve at peak is more suggestive of diffusion than shock interaction.

Other options require either a thermonuclear explosion, i.e., a pair- (PISN) or pulsational pair-instability supernova (PPISN), or rapid rotation and a central engine. Unmodified PPISNe are too low in energy to explain SN~2020wnt; PISNe require the retention of a large amount of mass;  and
rapid rotation is damped by mass loss. This means that successful models will be favored by low mass loss, and presumably low metallicity \citep[e.g.,][]{Woo06}, or require special circumstances, like a binary merger \citep[e.g.,][]{deM13,Mar19}. 

If attributed to diffusion, the broad main peak requires a large mass of H-poor ejecta, well in excess of 10 \Msun, with characteristic speeds in excess of 6000 km s$^{-1}$ (from nebular spectra). 
Taken together, this implies an explosion energy in excess of $\sim5 \times 10^{51}$ erg. 
Besides the fact that no neutrino-powered model has ever achieved such a high terminal energy, non-rotating stars with helium core masses this large are expected to collapse entirely to black holes \citep{Ert20}. 
We thus expect, from basic principles, that any successful model for SN\,2020wnt is going to involve rotation and be relatively ``non-standard''.  
The initial 5 -- 10 day spike attributed to ``break-out'' also requires its own explanation. 
True shock breakout from a compact Wolf-Rayet star generates a faint, very brief transient in the soft x-ray band \citep{Tol13}, quite unlike what was observed here.

Here we present models (Table.~\ref{tab:new_models}) computed using the \texttt{KEPLER} code \citep{Wea78,Woo02,Woo17}. For three PISN models, C95, C100, and C105 carbon-oxygen cores were evolved without mass loss through their stable carbon and neon burning phases and explosion. Due to the nearly fully convective nature of very massive stars during helium burning, even mild mass loss will uncover the CO-core in a star that has lost its hydrogen envelope, so pure CO-stars are a reasonable starting point.  For a lighter model, He20, a 20 \Msun \ helium star was evolved until iron-core collapse including mass loss according to \citet{Yoo17}. Two others, T50 and T70, assumed rapid rotation that led to chemically homogeneous evolution on the hydrogen-burning main sequence. These stars were evolved until iron-core collapse including mass loss according to \citet{Nie90} and \citet{Yoo17}. Except for the PISN, all pre-SN models were exploded using a piston situated near the edge of their iron cores and moved so as to generate a desired terminal kinetic energy. 
SN light curve calculations assumed an opacity due to electron scattering plus a background opacity due to lines of 0.03 $\rm cm^2 \, g^{-1}$.
We recognize the extreme uncertainty in the mass-loss rate in very massive stars, and emphasize that the relevant stellar parameters are those at the time of core collapse. 

Figure~\ref{fig:bolometric_LC} (right) and \ref{fig:LC_magnetar} compare the bolometric light curves of these models to the observed quasi-bolometric light curve. %

\subsection{A Poor Match with Pair-Instability Supernovae}
Figure~\ref{fig:bolometric_LC} (right) shows the bolometric light curve of SN\,2020wnt in comparison with PISN models, both from the literature and newly computed. 
We plot a bolometric light curve of the 100 $M_\odot$ H-poor PISN model from \cite{kasen2011}, along with the three new models of PISN progenitor stars based on 95, 100, and 105 \Msun\ carbon (10\%) and oxygen (90\%) cores.
These are models C95, C100, and C105 from Table~\ref{tab:new_models}.
In PISNe, the explosion energy, \nickel\ mass, and total mass are all inextricably entwined and cannot be varied independently. If the peak is powered entirely by radioactivity, then we determined the \nickel\ mass to be about 5 to 10 \Msun.
For a PISN, this requires a core mass close to 100 \Msun, much more than what we derived from the rise time of SN\,2020wnt.
The three PISN models illustrate this issue more concretely: models that are bright enough (C105) peak too late and models that peak early enough (C100) are too faint.
Though the \nickel\ mass and expansion speed of Model C100 is in reasonable accord with observations, its long rise time and broad peak are not. 
Further, the explosion time of SN\,2020wnt is well determined, so one cannot arbitrarily shift the time of peak to better fit the data. 
The introduction of metallicity \citep{kozyreva2017}, mixing \citep{kozyreva2015, gilmer2017}, multi-dimensional calculation \citep{chen2014}, and rotation \citep{chatzopoulos2013, chatzopoulos2015} affect the predicted observable properties of PISNe, but not enough to match our observations of SN\,2020wnt. 
Thus, we disfavor a classical PISN as the explosion mechanism for SN\,2020wnt. 

\subsection{Models That Include a Magnetar and Radioactivity}
With the only explosion mechanism known to produce a sufficient amount of \nickel\ to power the peak of SN\,2020wnt ruled out, we consider an additional power source from a central engine.
Though black hole accretion might be involved, the central engine is more easily parametrized as a magnetar. 
A constraint from the spectroscopic observations is that the central engine must be embedded and invisible during the optically thick phase to explain the lack of its spectroscopic signatures.
Thus, the central engine's energy output must be significantly smaller than the explosion energy, resulting in only the innermost ejecta directly affected by its heating.
An energetic explosion is still required to give the requisite combination of high velocity and a long diffusion time. 
The successful model will thus almost certainly make appreciable \nickel\ as well.
In general, models we consider have the rise time set by the ejecta mass and the explosion energy; the peak luminosity is set by the magnetar rotational energy; and the late-time luminosity set by the magnetic field of the magnetar.

These models resemble those of \citet{dessart2019}, but differ in focusing on more massive explosions with longer rise times. 
Some of our models also include rotation and interaction with a CSM. 
However, the radiation transport and spectra are better treated in \citet{dessart2019}.

\subsubsection{Lower Mass Non-Rotating Helium Star with Magnetar and CSM Interaction}
\label{magnet}

Model He20 (Table~\ref{tab:new_models}) is a non-rotating 20 \Msun\,helium star with initial metallicity 10\% solar that evolves including mass loss according to \citet{Yoo17}. It resembles Model r0e2 of \citet{dessart2019}, but has a larger ejected mass, 14.9~\Msun \ vs 9.7~\Msun, and therefore a longer rise time.
An artificial explosion was induced using a piston located at 2.00 \Msun, which was moved sufficiently rapidly to eject all external matter with high speed. 
The asymptotic kinetic energy of the ejecta was $4.3 \times 10^{51}$ erg. 
Because of the deeply situated piston and shallow density gradient outside, substantial $^{56}$Ni was synthesized, 0.88 \Msun. 
This large kinetic energy and nickel mass are inconsistent with a neutrino-powered explosion \citep{Ert20}, so formation of a rapidly rotating pulsar or accretion onto a black hole is implicitly assumed, even though rotation was neglected in the evolution. The model resembles those that have been proposed for gamma-ray bursts \citep[e.g.,][]{Woo06b,Mac99}, but does not involve the escape of a relativistic jet.

The light curve of the unmodified, radioactive model peaks below $10^{43} \ \rm erg\, s^{-1}$, far too faint to explain SN\,2020wnt.
Indeed, producing the peak of the light curve using radioactivity alone would require almost an order of magnitude more \nickel, about 7 \Msun. 
This would require a large explosion energy, which would greatly reduce the diffusion time and exclude models of such low mass. 
It would also require more \nickel\ than is physically credible for the density gradient surrounding the iron core in Model He20. 
Raising the explosion energy in this model to the extreme $3.5 \times 10^{52}$ only increases the \nickel\ production to 1.22 \Msun. 
Such a large energy would require a massive neutron star \citep{Met15}.
The pulsar would also not develop its full energy for a few seconds after collapse while the proto-neutron star cooled by neutrino emission. 
During that time, additional material would accrete on the proto-neutron star.

The effect of adding a magnetar was thus explored. 
An initial rotation rate of 4 ms, corresponding to a rotational energy of $1.2 \times 10^{51}$ erg, was assumed, and a constant magnetic field of $2.0 \times 10^{14}$ G. 
The energy deposition formula of \cite{woosley2010} was used, which differs slightly from \citet{kasen2010}. 
Energy was uniformly deposited in the inner 100 zones of ejecta (0.54 \Msun). 
This energy both illuminates the SN and inflates a low density bubble bounded by a high density shell out of which the trapped energy diffuses.
It is important that the magnetar energy, $1.2 \times 10^{51} \ \rm erg$, is considerably less than the background SN kinetic energy, $4.3 \times 10^{51} \ \rm erg$.
Consequently, in our 1D study, the bubble of pulsar heated energy does not erupt outside of the photosphere before the light curve peaks.
It remains bounded within the inner 7.62 \Msun\,of the 14.9 \Msun\,(51\%) ejecta and ultimately all moves at a nearly constant speed of 4400 km~s$^{-1}$.
In reality, multi-dimensional effects would lead to the broadening of the high density bounding the bubble by of order 10--29\% in radius \citep{Che16} and the swept-up matter would not all move at constant velocity.
This level of thickening would not lead to fingers of plasma extending above the photosphere until well after peak and this might have consequences for the spectroscopic history \citep{dessart2019, Kas16}.  
The model light
curve peaked at 65 days with a luminosity $4.7 \times 10^{43}$ erg s$^{-1}$. 
he photosphere reached the edge of the magnetar-inflated bubble about 80 days later, but this is a very approximate number given multi-dimensional effects and the simplifying assumptions made in \texttt{KEPLER} for radiation transport (e.g., spherical symmetry, single temperature, and simple opacity assumptions).

As Fig.~\ref{fig:LC_magnetar} shows, radioactivity plus an embedded magnetar replicates the broad features of the bolometric light curve near peak and on the tail, with the bulk of the energy being provided by the magnetar. Full trapping of magnetar-deposited energy is assumed, even though that is probably unlikely at the latest times plotted. 
The light curve shown is thus an upper bound for the magnetar contribution. 
On the other hand, even a slight decrease in the magnetic field would give a light curve that was brighter at late times.

The magnetar plus radioactivity model alone still fails to account for the bright transient lasting about a week at the beginning of the SN. 
That feature is almost certainly due to CSM interaction.  
The outer 0.02 \Msun\,of Model He20 has a speed in excess of 20,000 km s$^{-1}$ and carries $1.1 \times 10^{50}$ erg, more than enough to explain the initial display, provided that ejecta interacted with a comparable mass inside $\sim10^{16}$ cm. 
Figure~\ref{fig:LC_magnetar} includes the effect of adding pre-explosive mass loss around the SN. 
A high mass-loss rate is necessary to explain the initial bright peak, but the same mass loss would over-illuminate the dip at $\sim20$ d. 
One must therefore make the reasonable assumption that the mass loss increased as the star neared its death. 
We have in mind the acoustic transport of convective energy during the post-carbon burning stages as discussed by \citet{Shi14, fuller2018, leung2021}. 
Here a mass-loss rate of 0.0015 \Msun\,yr$^{-1}$ was adopted during years 3.75 to 0.75 before iron core collapse, rising to 0.01 \Msun\,yr$^{-1}$ in the final 0.75 years, for a total CSM mass of 0.012 \Msun.  
This CSM mass is roughly consistent with what is predicted by \citet{leung2021}.
A steady wind speed of 500 km s$^{-1}$ is assumed, but other values of mass loss and wind speed would also give similar results, provided that $\dot M/v_{\rm wind} \sim 10^{15}$ -- 10$^{16}$ ~g~cm$^{-1}$ during the last few years. A still smaller mass-loss rate at earlier times might characterize the tail of the light curve.

\subsubsection{A More Massive Magnetar Model With Rotation}\label{sec:magrot}

Now consider a more massive model evolved from the main sequence including mass loss and rapid rotation. 
Similar models have been considered for the progenitors of gamma-ray bursts and SLSNe~Ic \citep{aguilera2018, Woo06b}.
Model T50 begins its life as a 50 \Msun\,main sequence star with 10\% solar metallicity and an equatorial rotational speed of 240 km s$^{-1}$. 
The initial angular momentum of the star on the ZAMS was $1.5 \times 10^{53}$ erg s. 
With this large rotation rate, this star evolved chemically homogeneously, burning most of its hydrogen on the main sequence and avoiding red giant formation. It thus avoids the large shedding of mass and angular momentum that would occur in that phase. 
Using the \citet{Yoo17} mass-loss rate appropriate for 10\% solar metallicity after central hydrogen depletion, the pre-SN star had a mass of 28.36 \Msun\,and an iron core of 2.41 \Msun. 
There was a sharp fall off in density at the base of the oxygen shell at 2.50 \Msun. 
The matter interior to 2.50 \Msun\,had an angular momentum of $2.4 \times 10^{49}$ erg s. 
If it collapsed to a neutron star, that compact remnant would have a gravitational mass of about 2.1 \Msun\,and a sub-millisecond rotation period.

While the subsequent evolution of such a core is uncertain, it is plausible that some fraction of the available several $\times 10^{52}$ erg rotational energy of a massive, sub-ms pulsar \citep{Met15} will be released in about a second.  
Accordingly, a piston was situated at 2.50 \Msun\,and moved so as to produce an asymptotic kinetic energy $1.37 \times 10^{52}$ erg. 
The actual work done by the piston was larger, $1.67 \times 10^{52} \ \rm erg$, since the net binding energy of the pre-SN star external to 2.50 \Msun was $3.05 \times 10^{51}$ erg. 
The explosion also produced 0.85 \Msun\,of \nickel. 
If the piston had been situated deeper, this \nickel\ mass might be increased by at most a few tenths \Msun, but not much more.

Even with a higher \nickel\ mass, the resulting light curve would still fail to match that of SN\,2020wnt. 
To agree with observations, the light curve must again be supplemented with a central energy source to boost the peak luminosity, and CSM interaction to explain the initial peak. 
In order to boost the speed of the inner ejecta and provide a slightly greater luminosity at peak, a magnetar of slightly greater initial rotational energy,
$2 \times 10^{51}$ erg (P = 3 ms), was used here instead of the slower rotating $1.2 \times 10^{51}$ erg used for Model He20. 
The magnetic field, $2.0 \times 10^{14}$ G, which sets the late time luminosity, was the same. 
Because of its greater energy, the model was also surrounded by a slightly more massive circumstellar medium, 0.015 \Msun, again consisting of two shells, 0.001 \Msun yr$^{-1}$ for years 6 to 1 before explosion and 0.01 \Msun\,yr$^{-1}$ for the last year. 
The resulting light curve is shown in Fig.~\ref{fig:LC_magnetar}. 
The final kinetic energy of the ejecta in the magnetar model is $1.50 \times 10^{52}$ erg which includes $1.37 \times 10^{52}$ erg of net energy
from the piston, $2 \times 10^{51}$~erg for the magnetar, less $0.7 \times 10^{51}$ erg radiated as light. 
Due to the action of the magnetar, the inner 10.6 \Msun\ of the 25.8 \Msun\ of ejecta (41\%) was accelerated into a thin shell moving at 5700 $\rm km \, s^{-1}$.
The model had a peak luminosity of $6.6\times 10^{43} \ \rm erg\, s^{-1}$ 60 days after the explosion. 
The magnetar-inflated bubble was still well below the photosphere at that time.

\subsubsection{Pulsational Pair-Instability Model Plus Central Engine}\label{sec:ppisn}

Pulling out all the stops, we considered the most massive credible model for SN 2020wnt, a PPISN plus magnetar. 
A more massive model would either eject so much mass during its pulsations that a very different light curve would result \citep{Woo17}, or explode completely as a PISN. 
For the assumed reaction rates, and hence central carbon mass fraction at helium exhaustion, the pair instability is first encountered for pre-SN core masses $\gtrsim35$ \Msun. 
Though often thought of as a violent instability ejecting solar masses of material at a time, on its lighter end, in the less massive, presumably more abundant stars, the pair instability only causes a series of weak pulses that cumulatively eject a few solar masses or less at moderate speeds (few 1000 km s$^{-1}$) over a period of a several days. 
The remaining star then collapses to a black hole or neutron star 
(see for example the 38 \Msun\,model in Table 1 of \citealt{Woo17}). 
If all the remaining star did was collapse to a black hole, this sort of model would only produce a very faint, brief transient coming from the small internal energy of the ejected helium and heavy elements. 
If, however, iron core collapse somehow leads to an energetic explosion, the interaction with several solar masses of ejecta has interesting implications for the early light curve.

Model T70 (Table~\ref{tab:new_models}) began its life as a 70 \Msun, rapidly rotating main sequence star with 10\% solar metallicity similar to
Model T50, but with greater mass and angular momentum, $3.5 \times 10^{53}$ erg s.
The equatorial rotation speed was 380 km s$^{-1}$, or about 20\% the Keplerian speed. 
The star again experienced CHE and mass loss, ending its life as a rapidly rotating core of mostly carbon, oxygen and heavier elements. 
For the pre-SN star, the mass interior to the mass cut at 2.50 \Msun\, had angular momentum $3.9 \times 10^{49}$ erg s, sufficient to make a sub-ms pulsar with gravitational mass roughly 2.1 \Msun\, if black hole formation were avoided. 
Indeed the inferred rotation rate would be about 0.4 ms, suggesting rotation would play a large role in the iron core's subsequent contraction and evolution.

Before dying, the star became unstable to nuclear energized pair-instability pulsations in the oxygen burning shell after a silicon burning core had already been established. 
Dozens of weak pulses ejected a total of 3.2 \Msun\,with a typical speed of 2000 -- 4000 km s$^{-1}$ and a total energy of $2.3 \times 10^{50}$ erg. 
By the time the core collapsed this matter had coasted to radii $\sim10^{13} - 10^{14}$ cm, forming something that resembled a red supergiant envelope around the collapsing core.

An explosion was simulated with a piston at 2.50 \Msun\,with a terminal kinetic energy of $1.76 \times 10^{52}$ erg, plus binding energy, $4.8\times 10^{51}$ erg, for a total of $2.24 \times 10^{52}$ erg. 
Because of the deep mass cut, even slightly into the iron core itself, the shallow density gradient outside this core, and a very energetic explosion, the model ejected 1.8 \Msun\,of \nickel, which we regard as close to an upper bound. 
Raising the explosion energy to $3.5 \times 10^{52}$ erg only increased the $^{56}$Ni mass to 2.25 \Msun. 
The blast also interacted with the 3.2 \Msun\,of slow moving ejecta from the prior pulsations producing the "break-out" transient. 
The sharp dip at about 20 days would have been brighter if the pulsations had ejected their matter just a bit sooner or if there were a wind of only 0.001 \Msun\,yr$^{-1}$ prior to the pulsations.

Like previous models, even this extremely massive model cannot produce SN\,2020wnt's light curve without a central engine. 
The model light curve in Fig.~\ref{fig:LC_magnetar} shows the effect of adding a magnetar with initial rotational energy $2 \times 10^{51}$ erg and magnetic field $2.0 \times 10^{14}$ G.
The light curve peaks at 63 days with a luminosity of $8.1 \times 10^{43} \ \rm erg \, s^{-1}$.
The magnetar accelerates the inner 13.5 \Msun\ of the 39.3 \Msun ejecta (34\%) to a constant speed of 5100 $\rm km \, s^{-1}$.
Similar to other magnetar models, the effects of the magnetar are well hidden at peak.

\subsection{Best-Fit Model for SN 2020wnt}
Three promising models for SN 2020wnt involving similar central engines have been  considered. 
Magnetars are responsible for producing the peak of the light curve in all of them, and CSM interaction produces the bright initial transient.  
For the two models with winds, He20 and T50, the rapid decline and sustained emission at around 20 days implies a mass-loss rate that increased just before the star died.
The required loss rates, 0.01 to 0.001 \Msun \ yr$^{-1}$ during the last several years, are consistent with predictions by \citet{Shi14, leung2021}. 

The maximum $^{56}$Ni mass seen in any calculation, even the most massive possible model with the deepest mass cut and an explosion energy near $4 \times 10^{52}$ erg, was 2.25 \Msun. 
This is inadequate to explain the light curve peak, though such a large amount of \nickel\ would still have consequences for the spectrum and for the light curve on the tail. None of the simple magnetar-powered models reproduced the dip seen on the tail of the light curve, which is likely due to dust formation.

\texttt{KEPLER} does not compute spectra, yet from the small interaction radius, $\sim10^{14}$ cm, and high luminosity, $\gtrsim 10^{43}$ erg s$^{-1}$ of the break out transient, we can confidently say that the model spectrum at that stage would have been very blue. 
Indeed, much of the light during the breakout transient may have been emitted in the far ultraviolet, and we only observed the tail of the black body radiation. 
The observed $g-r$ color was indeed blue (Fig.~\ref{fig:photometry}), indicating a large bolometric correction. 
Closer to the main peak, typical effective temperatures were $\sim8,000$ K. 
The photosphere at peak was still in the outer part of the ejecta, outside of the magnetar-inflated bubble. 
In that case, it may be difficult to distinguish, even spectroscopically, between an embedded magnetar and radioactivity as the primary power source.

While we have emphasized a magnetar energy source for the peak luminosity, black hole accretion from fallback is an interesting alternative if a neutron star did not form \citep{Mac01, dexter2013,  moriya2019}.
This might be expected in such massive progenitor stars.
The time history of this accretion is difficult to predict, however, given the uncertain angular momentum distribution in the ejecta and the role of the reverse shock \citep{dexter2013}.

\section{Spectra Modeling}\label{sec:nebular_spec_modeling}

\subsection{Nebular Spectrum Comparison}
In the nebular phase, the ejecta are optically thin, which allows our observations to measure their composition and constrain their properties, including the total mass. 
We compare our optical to near-IR spectra from 301 to 313 d post-peak (about 378 d post-explosion) to the single-zone model grid from \cite{jerkstrand2017}, computed using the \texttt{SUMO} code.
To prepare our data for this comparison, we perform absolute flux calibration of the optical spectrum by matching it to the flux observed in the $r$ band at that epoch.
We then scale the near-IR spectrum to match the optical in the overlapping region. 
The models are computed at 400 d post-explosion assuming the ejecta velocity of $8000 \ \rm km \, s^{-1}$ to set the physical volume of the ejecta containing the mass $M_{\rm ej}$. 
While these numbers are not exactly the same as the epoch of observation and ejecta velocity of SN\,2020wnt, the grid is too coarse for these small differences to matter. 
The models assume 100 spherical clumps of ejecta within this volume, taking up a filling factor of $f$. 
This factor sets the density within each clump: smaller $f$ results in a larger density.  
The models account for only the O-rich zone of the ejecta, which is responsible for most of the emission lines observed at these epochs.
By not including other subdominant zones, these models are less computationally intensive and are ideal for parameter space exploration. 
We only consider full carbon-burning composition models from the paper (and not the pure oxygen or the oxygen-magnesium models). 
The ejecta composition in these models is 74\% O, 15\% Ne and 7\% Mg. 
The model grids we use are computed over the ejecta masses of 3, 10, and 30 $M_\odot$ (only O-rich zone); the filling factors of 0.1, 0.01, and 0.001; and the powering luminosities of 2.5, 5, 10, and 20$\times 10^{41} \ \rm erg\,s^{-1}$. 
The model is agnostic to the nature of the source of the powering luminosity, though the paper provides discussions on how clumpy ejecta could be indicative of a central engine.
From our bolometric light curve, the luminosity at the epoch of the spectra is about $10^{42} \ \rm erg\, s^{-1}$.

\begin{figure*}[!ht]
\centering
    \includegraphics[width = \linewidth]{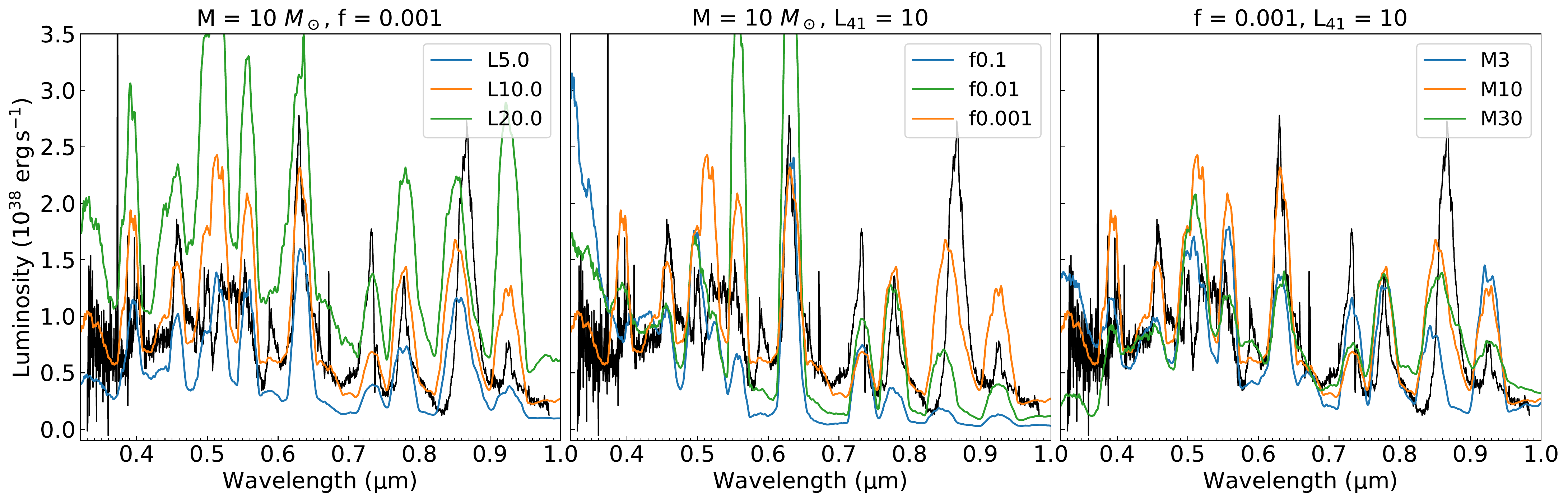} \hfill
    \includegraphics[width = \linewidth]{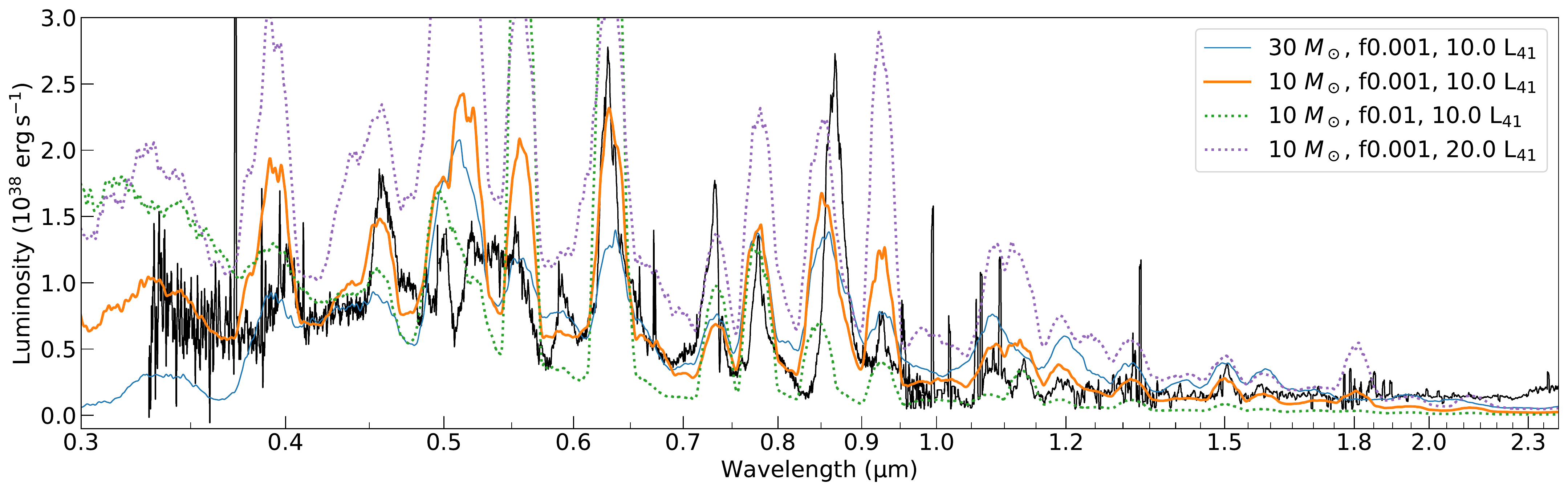}
    \caption{Comparisons between a nebular-phase spectrum of SN\,2020wnt and models from \protect\cite{jerkstrand2017}. 
    \textbf{Top row} compares the optical part of the spectrum with models with varying parameters. 
    The model names, e.g., M10 f0.001 L10, refers to the model with 10 $M_\odot$ of ejecta mass, a filling factor of 0.001 and a energy deposition rate of $10^{42} \ \rm erg \, s^{-1}$. 
    Similarly, $\rm L_{41}$ refers to the unit of $10^{41} \ \rm erg\, s^{-1}$.
    Panels from left to right show models with varying energy deposition rate, filling factor, and mass respectively. 
    The best-fit model, M10 f0.001 L10, is plotted with the same color in all panels. 
    \textbf{Bottom} shows a similar comparison, but with the entire optical-NIR spectrum. Note that the $x$ axis is in a logarithmic scale.
    Four models are shown here for comparison.
    The same best-fit model, M10 f0.001 L10, is again plotted in the same color as in the top panels.
    }
    \label{fig:nebular_model_comparison}
\end{figure*}

Figure~\ref{fig:nebular_model_comparison} shows select results from this comparison.
The top three panels compare the optical part of the spectrum to models with varying powering luminosity, filling factor, and mass, respectively. 
The relationship between the luminosity and the resulting spectra is straightforward: larger powering luminosity results in larger observed luminosity with modest effect on the lines shape and ratios.
Shown in the top left panel, we find that models with $L = 10^{42} \ \rm erg\, s^{-1}$, similar to the bolometric luminosity at this epoch, provide the best fit to the data.  

The next parameter we consider is the filling factor, which sets the density of the ejecta clumps (top middle panel).
Consequently, this parameter strongly affects the temperature and the line ratio.
The $f = 0.1$ model is the least dense and thus gets very hot, over predicting the flux in the blue while under predicting the red continuum and the line fluxes. 
The $f = 0.01$ model does better at matching the \ion{O}{1}~7774~\AA\ line, but is still over predicting the blue flux and other line fluxes, primarily the [\ion{O}{1}] 6300, 6363 \AA\ lines. 
Lastly, the clumpiest $f = 0.001$ provides the best fit to the data by simultaneously fitting the continuum across the optical band, and providing a decent fit to both the [\ion{O}{1}] 6300, 6363 \AA\ and \ion{O}{1}~7774~\AA\ lines. 
It is also the only model that fits the \ion{Mg}{1}] 4571 \AA\ line. 

Lastly, we consider different ejecta masses in the right panel. 
The differences between the three masses considered in this model grid are relatively subtle, with the [\ion{O}{1}] 6300, 6363 \AA\ lines being the most sensitive mass indicator. 
Our observation is best explained here by the moderate 10 $M_\odot$ model, with the other two masses producing poorer fits to the O lines, including in the near-IR. 
Further, it is the only model that can satisfactorily fit the \ion{Mg}{1}] 4571 \AA\ and \ion{Mg}{1} 1.5 $\mu$m lines.

We note, however, that some line predictions are still off in our best-fit model.
For instance, the [\ion{Ca}{2}] 7292 \AA\ and the Ca triplet are poorly fit.
The region between 5000 to 6000 \AA, which has many Fe lines, is also poorly fit. 
These discrepancies are most likely because features from these species do not come from the O-rich zone considered in these models. 
The ratio between the \ion{O}{1}~7774~\AA\ and 9263~\AA\ lines are different from the observed ratio, and the modeled line width is also larger than what is observed. 
As discussed by \cite{nicholl2019}, this could be evidence that the central engine is ionizing the innermost ejecta (smallest velocity), producing these narrow O recombination lines.
The presence of a central engine would also be consistent with the highly clumpy ejecta (small filling factor) that we infer. 
Further investigation using multi-zone ejecta models will produce a more realistic map the massive ejecta of SN\,2020wnt.

\subsection{Spectroscopic Comparison with Massive Core Collapse and Pair-Instability Models}\label{subsec:nir_nebular_comparison}
\begin{figure}
    \centering
    \includegraphics[width=\linewidth]{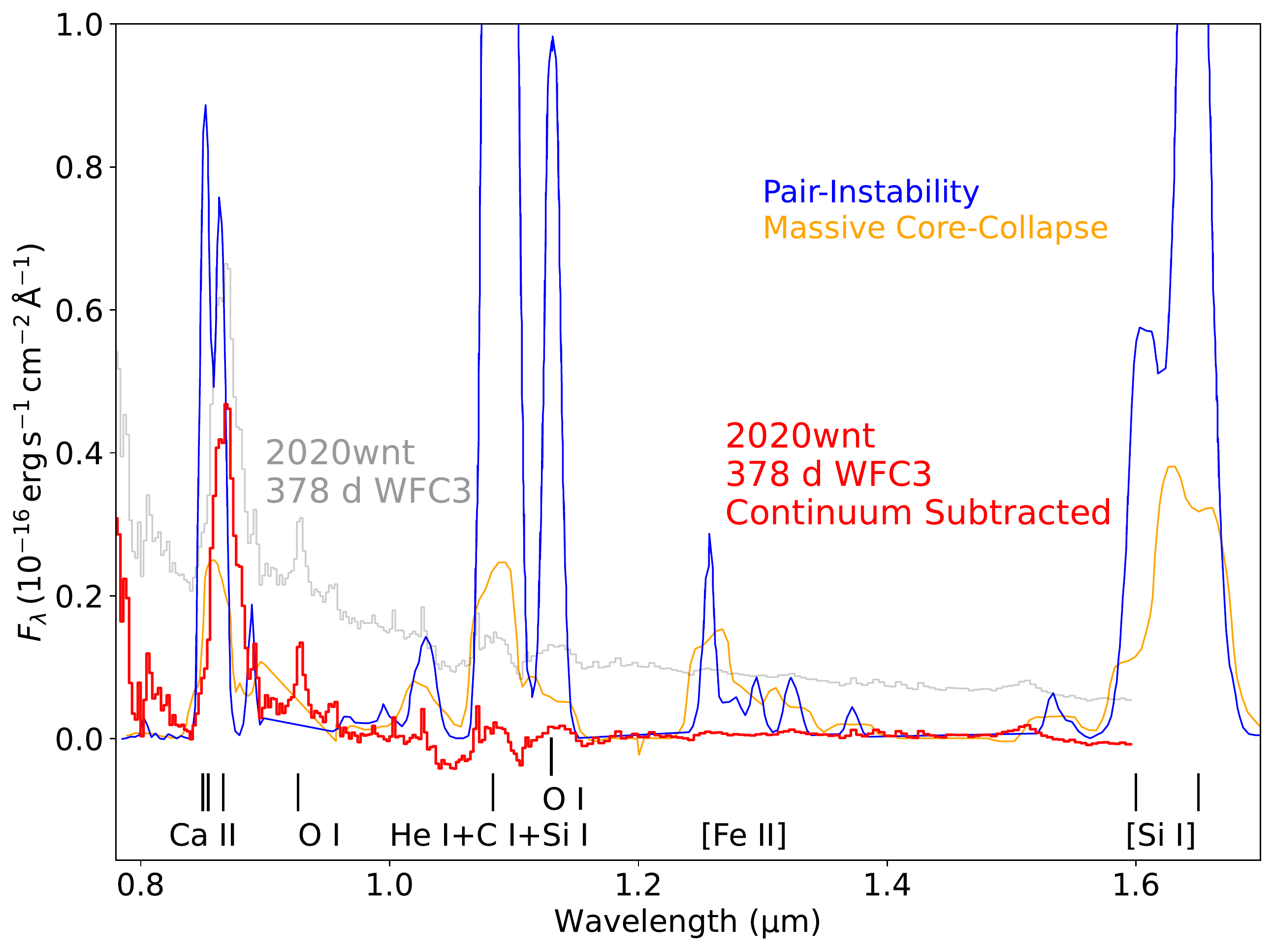}
    \caption{A comparison between the near-IR spectrum of SN\,2020wnt obtained at 378 d post peak with \textit{HST}/WFC3 near-IR grisms, and models of nebular spectra from a massive CCSN (orange) and a PISN (blue) from \protect\cite{mazzali2019}. 
    The models have been scaled to the distance of SN\,2020wnt. 
    The grey line shows the observed spectrum, which clearly has some continuum.
    While the underlying host may contribute, the P-Cygni absorption associated with the 1.08~$\mu$m feature shows that some of the continuum is intrinsic to the SN. 
    We estimate and subtract the continuum by fitting a low-order polynomial.
    The continuum-subtracted spectrum is shown in red. 
    }
    \label{fig:nebular_IR_model_comparison}
\end{figure}

Figure~\ref{fig:nebular_IR_model_comparison} shows a comparison between the \textit{HST}/WFC3 spectrum of SN\,2020wnt at 378 d post peak and the models of massive CCSNe and PISNe from \cite{mazzali2019}.
The observed spectrum plotted in grey shows a clear continuum still at this epoch. 
Some contribution to this continuum comes from the underlying \ion{H}{2} regions in the host galaxy.
However, we note that there is P-Cygni absorption associated with the 1.08~$\mu$m line, which demonstrates that some of the continuum is intrinsic to the SN. 

The model spectra from \cite{mazzali2019} were computed to fit the observations of SN\,2007bi at about 367 d post peak.
Here we scaled the models to the distance of SN\,2020wnt (141.8 Mpc from SN\,2007bi's 592 Mpc used in the paper). 
The massive CCSN model assumes 33 $M_\odot$ of ejecta mass with an explosion energy of $3 \times 10^{52} \, \rm erg$. 
The explosion produces 4.5 $M_\odot$ of \nickel. 
These parameters are in agreement with what we derived for SN\,2020wnt, if we assume that the SN is powered by radioactivity (note the caveats from \textsection\ref{sec:model_comparison}).
A core-collapse explosion with such an energy does not come from a normal neutrino-driven explosion. %
The PISN model shown here is from the 100 $M_\odot$ model with an ejecta mass of $94 \  M_\odot$ and a \nickel\ mass of 3.1 $M_\odot$. 
The main distinguishing features in the near-IR between these two classes of models are the strengths of the [\ion{Si}{1}] lines around 1.60 and 1.65 $\mu$m, and the line complex at 1.08 $\mu$m. 

In comparison to the models, SN\,2020wnt at 378 d post peak does not show any of the strong forbidden lines predicted by the PISN model. 
Specifically, the PISN model predicts strong lines around 1.08 $\mu$m from [\ion{S}{1}] and the [\ion{Fe}{2}] complex around 1.25 $\mu$m, in contrast with the observations. 
In general, the line strengths observed in SN\,2020wnt is much closer to those predicted by the massive core-collapse model.
One major exception is the [\ion{Fe}{2}] complex around 1.25 $\mu$m, also predicted by the core collapse model, is still missing. 
This could be because the spectrum of SN\,2020wnt does not appear to be completely nebular yet, and a later NIR spectrum would be crucial to make a more robust model comparison. 
However, the lack of distinct PISN features is another argument against SN\,2020wnt originating from a pair-instability explosion.

\section{Host Properties}\label{sec:host}

WISEA J034638.04+431348.3 is the host galaxy of SN\,2020wnt. 
It is a faint dwarf galaxy, with irregular morphology and no prior spectroscopic redshift. 
To measure the host galaxy redshift, we use the Keck/LRIS spectrum obtained on 2022 September 22. 
At this time, the SN was still visible in the spectrum, so the slit was positioned to pass through the host galaxy nucleus and we extracted the spectrum at this location. 
We derive a redshift for the host galaxy (based on the H$\alpha$ emission line) of $z=0.0323 \pm 0.0001$ and a distance of 141.8 Mpc, in agreement with \citep{gutierrez2022}. 
SN\,2020wnt has a radial offset of 6\farcs3 from the host galaxy nucleus, corresponding to a physical offset of 4.3~kpc at the distance of the host.

To derive host galaxy properties of SN\,2020wnt, we first measure the broadband photometric properties. 
Due to the irregular nature of the galaxy, we carefully perform elliptical aperture photometry (radius of 7\farcs9, axis ratio of 0.6 and angle 30 degrees east of north) on PS1 images \citep{Chambers2016}.  
Host photometry is detailed in Table \ref{tab:hostgalaxyphotometry}.
We note that our measured photometry differ from that in the PS1 catalog due primarily to the consistent aperture sizes used in different bands. 

We model the host galaxy SED based on our PS1 photometry using the \textsc{Le PHARE} package \citep{Arnouts1999MNRAS, Ilbert2006}, correcting the photometry for Milky Way foreground extinction prior to fitting. The code uses the population-synthesis templates of \citet{Bruzual2003MNRAS}, summed according to an exponentially declining burst of star formation and with stellar metallicities between $0.2 Z_{\odot} < Z < Z_{\odot}$ and assuming a Chabrier IMF \citep{Chabrier2003}.  Dust attenuation in the galaxy is applied to the SED models using the \citet{Calzetti2000} reddening law. We derive a host stellar mass of log$(M/M_\odot)$~=~$8.81^{+0.09}_{-0.18}$ and an integrated SFR of $\log({\rm SFR}) = -1.58^{+0.31}_{-0.40}$~M$_{\odot}$~yr$^{-1}$, respectively.  The values represent best fit SED and 1-$\sigma$ uncertainties derived from MCMC simulations.

This is 0.51 dex higher in stellar mass than in \citep{gutierrez2022}. 
However, the photometry used in \citep{gutierrez2022} was the catalog PS photometry (4\farcs4 Kron radius for the $r$-band, centered on the nuclear region of the host), which does not encompass all the flux of the galaxy, given the irregular morphology. 
When we fit a model SED to this catalog PS1 photometry we derive a a host stellar mass of log$(M/M_\odot)$~=~$8.30^{+0.10}_{-0.11}$ and an integrated SFR of $\log({\rm SFR}) = -2.55^{+0.61}_{-0.38}$~M$_{\odot}$~yr$^{-1}$, which is consistent with the measurements in \citep{gutierrez2022}. 

\begin{figure*}
\includegraphics[width=\textwidth]{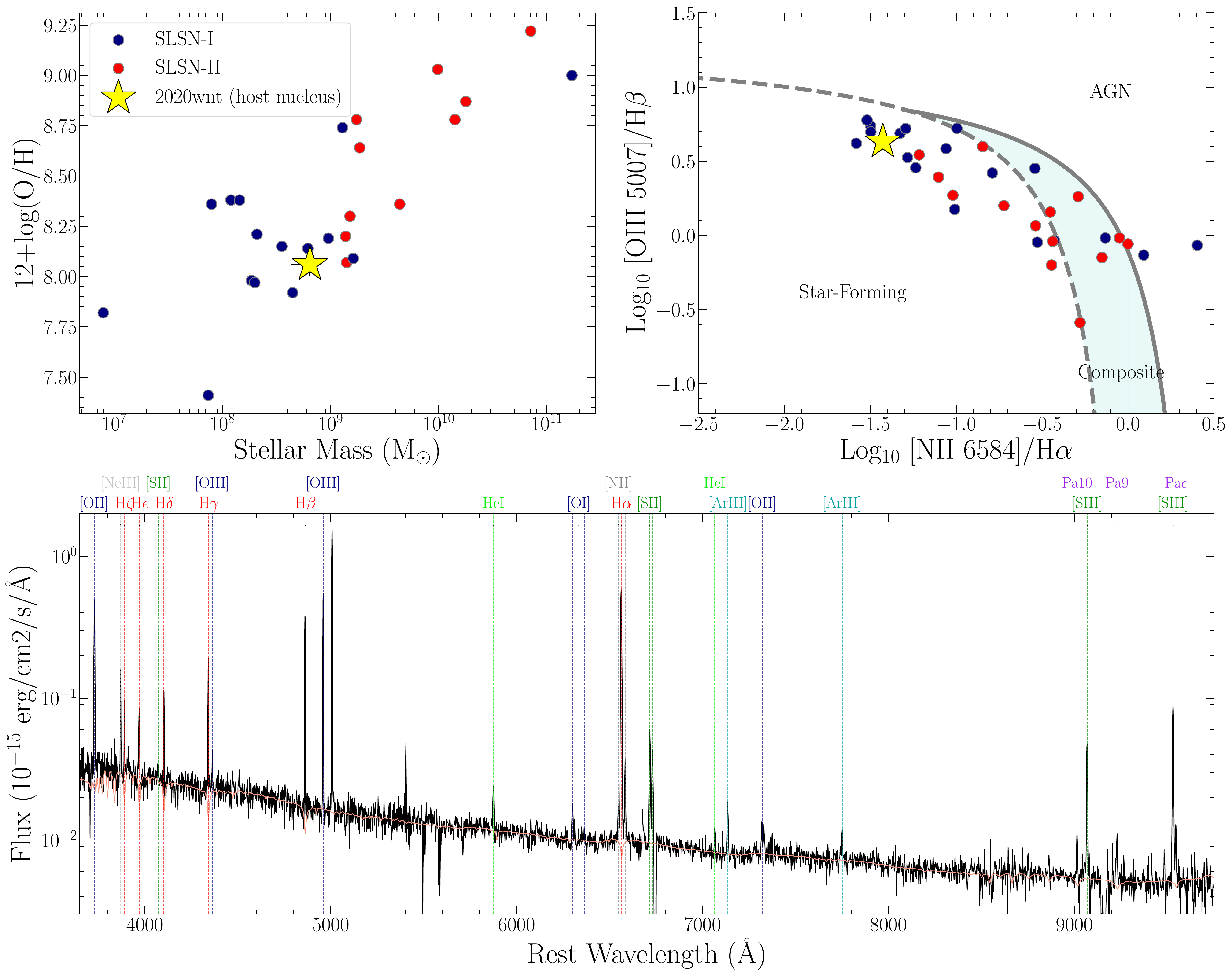}
    \caption{\textbf{Top left:} Stellar mass versus host galaxy metallicity of 2020wnt (yellow star) versus SLSNe-I (blue circles) and SLSNe-II (red circles) from \citet{Perley2016c}. \textbf{Top right:} BPT diagram showing 2020wnt is consistent with the star forming galaxy population. 
    \textbf{Bottom:} A Keck/LRIS spectrum of the host galaxy nucleus of 2020wnt.  Overplotted in tangerine is the best-fit SED model for the galaxy stellar continuum that is subtracted before measuring line fluxes. Strong emission lines are identified. Notably, the auroral oxygen lines of  [\ion{O}{3}] $\lambda$4363 and [\ion{O}{2}] $\lambda$7320,7330 are present allowing a direct metallicity measurement.}%
    \label{fig:2020wnt_galaxy_spec_properties}
\end{figure*}

We then constrain the spectroscopic properties from our Keck/LRIS spectrum.
We corrected the host galaxy for Milky Way reddening \citep{schlafly2011}, modelled and the host galaxy stellar continuum using the \texttt{Firefly} code\footnote{\url{https://github.com/FireflySpectra/firefly_release}} \citep{Wilkinson2017}
with MaStar Models \citep{Maraston2020}. 
The stellar continuum model is convolved to the resolution of the spectrum and subtracted to remove the stellar Balmer absorption. 
The Galactic reddening corrected spectrum is plotted in Figure~\ref{fig:2020wnt_galaxy_spec_properties}, and the stellar continuum is plotted in a tangerine color. Emission line features are labelled. 

We then measure emission-line fluxes from our one dimensional spectrum by fitting a Gaussian profile for each line of interest. 
Flux values are available in Table~\ref{tab:hostgalaxyemissionlines}. 
We calculate the Balmer decrement extinction ${\rm H}\alpha$/${\rm H}\beta$ to estimate the visual extinction as A$_V$=0.006$^{+0.173}_{-0.006}$ mag, assuming the \citet{Calzetti2000} attenuation law.

We plot SN~2020wnt on a BPT diagram \citep{Baldwin1981,Veilleux1987} to check for contamination with an AGN. In the top right panel of Fig. \ref{fig:2020wnt_galaxy_spec_properties} we plot SN~2020wnt alongside other SLSNe from \citet{Perley2016c}. SN~2020wnt clearly falls on the star-forming main sequence, and shows no evidence for harbouring an AGN. We then measured the spectroscopic SFR using the H$\alpha$ line. We corrected the H$\alpha$ flux for extinction and used the H$\alpha$ luminosity with the standard conversion of \citet{Kennicutt1998}, to derive a star-formation rate of $\log({\rm SFR_{H\alpha} }) = -1.12^{+0.07}_{-0.01}$~M$_{\odot}$~yr$^{-1}$, which is 0.45 dex higher than the photometric SFR measurement, but still consistent within the measurement uncertainties. 

In addition, we derived the nuclear metallicity of the host galaxy. We detected the temperature-sensitive [\ion{O}{2}]\,$\lambda$\,4363 auroral oxygen feature ($9.9~\sigma$), as well as [\ion{O}{2}]\,$\lambda$\,7720, 7730 (7.0,5.0~$\sigma$). Thus, employed the direct method to measure the metallicity. We used the python package \texttt{PyNeb version 1.1.16} \citet{Luridiana2015a}. We assumed a two-zone model for the \ion{H}{2} regions, where the high ionization zone is traced by [\ion{O}{3}] and the low ionization zone is traced by [\ion{O}{2}], [\ion{N}{2}] and [\ion{S}{2}] \citep[e.g.,][]{Stasinska1982}. To get from the observed line fluxes to the metallicity, we first corrected all line fluxes for the attenuation in the host galaxy. Afterwards, we computed the electron density from the ratio of [\ion{S}{2}] doublet and the electron temperature of the high ionisation species from the flux ratio between [\ion{O}{3}]\,$\lambda$\,4363 and [\ion{O}{3}]\,$\lambda$\,4959 + [\ion{O}{3}]\,$\lambda$\,5007. The temperature of the low-ionisation species is computed with the Eq. 14 in \citet{Izotov2006} for the intermediate metallicity case ($12+\log\,\rm O/H \approx 7.6$). With the electron temperatures in hand (Table \ref{tab:hostgalaxyemissionlines}), we can directly infer the oxygen abundance of O$^+$ and O$^{2+}$ with \texttt{PyNeb}: $12 + \log\,\rm O^+/H = 7.54^{+0.07}_{-0.06}$ and $12 + \log\,\rm O^{2+}/H = 7.91^{+0.06}_{-0.05}$ (Table \ref{tab:hostgalaxyemissionlines}). The total oxygen abundance is $12 + \log\,\rm O/H = 8.06\pm0.06$. This corresponds to metallicity of $0.24 \pm 0.03$ solar, using a solar oxygen abundance of 8.69 \citep{Asplund2009}. 

 The top left panel of Fig. \ref{fig:2020wnt_galaxy_spec_properties} shows a mass-metallicity diagram. SN~2020wnt is consistent with other SLSNe-I. It has a slightly lower metallicity (0.14 dex) than the average metallicity of SLSNe-I host galaxies \citep[$12+\log\,\rm O/H\approx8.2$;][]{Leloudas2015a,Perley2016c} and the stellar mass is higher (0.91 dex) than relative than the average stellar mass of other low-redshift SLSNe-I host galaxies \citep[log($M/M_\odot$)$\approx$7.9;][]{Perley2016c,Schulze2021,Taggart2021}.

\section{Summary}\label{sec:conclusion}
SN\,2020wnt presents one of the best views yet of a rare death of a very massive star. 
From the optical and near-IR observations we present in this paper, we find the following:
\begin{enumerate}
    \item The light curve of SN\,2020wnt consists of a $\sim$5-day long early-time bump, a diffusive peak, and a nebular decline with a 1-mag dip around 200 d post peak. 
    The early bump is likely caused by the presence of a compact CSM ejected immediately before the explosion, possibly by nuclear burning instability. 
    The shock cooling from the interactions between this CSM and the SN shock results in the emission we observe. 
    The diffusive peak appears similar to a light curve of SNe~Ibc and not that of SLSNe~I. 
    Lastly, the nebular decline is consistent with several models considered. 
    The dip is most likely due to dust formation. 
    The rise time of the bolometric light curve is $69 \pm 2$ d to the peak luminosity of $(6.8 \pm 0.3) \times 10^{43} \ \rm erg\, s^{-1}$.
    \item The optical spectra of SN\,2020wnt prior to 100 d post peak are quantitatively similar to those of SNe~Ic and not magnetar-powered SLSNe~I (Fig.~\ref{fig:type_score}, \ref{fig:spec_comparison}). 
    The blue continuum and \ion{O}{2} absorption lines, associated with magnetar heating, are not present. 
    However, late-time spectra much better resemble the mean SLSN~I late-time spectrum rather than the mean SN~Ic spectrum. 
    Spectroscopic signatures attributed to a central engine in SLSNe~I are visible in SN\,2020wnt's spectra at these epochs: enhanced blue flux, an emission feature at 5000 \AA\ associated with [\ion{Fe}{2}] and [\ion{O}{3}], and narrow \ion{O}{1} recombination lines. 
    This spectral evolution suggests that a central engine may be operating in SN\,2020wnt, but is not visible until later in the nebular phase. 
    \item The extensive near-IR spectroscopic data of SN\,2020wnt reveals features not captured in the optical. The most notable is the emerging CO emission at 2.3 $\mu$m starting at 241 d post peak and the rising dust continuum. 
    These observations support our conclusion that the dip in the optical light curves around 200 d post peak is due to additional dust absorption. 
    SN\,2020wnt is the first SLSN with a CO detection, thanks to its low redshift allowing for the observation of the 2.3 $\mu$m band from the ground. 
    It is also the second SLSN with an observation of dust continuum (after SN\,2018bsz, \citealp{chen2021}).
    This finding highlights the need for space-based IR observations of SLSNe with \textit{JWST}.
    \item Our light curve modeling effort disfavors a pair-instability SN as the explosion mechanism responsible for SN\,2020wnt. The predicted light curves evolve too slowly and it is not possible to reproduce both the rise time and the peak luminosity.
    The large \nickel\ mass that would be require to power the peak of SN\,2020wnt is simply too large to synthesize without a pair-instability explosion.
    We instead favor magnetar models, with a large enough total ejecta mass to hide the spectroscopic effects from magnetar heating until later in the evolution. 
    The T50 model from Table~\ref{tab:new_models} is our preferred model. 
    \item Our nebular spectra comparison also disfavors the pair-instability model (Fig~\ref{fig:nebular_IR_model_comparison}) due to the lack of strong intermediate-mass element lines predicted.
    The nebular spectra comparison to a more generic model grid shows that the ejecta of SN\,2020wnt are likely very clumpy, suggesting the presence of a central engine. 
\end{enumerate}
SN\,2020wnt demonstrates the rich and diverse observational properties of a magnetar-powered explosion. 
Dust and molecule formation remain poorly probed; it would require future \textit{JWST} observations of these objects, especially at low redshift, to create the first sample of SLSNe with rest-frame IR spectra and to map the chemical evolution of SLSN ejecta.
More importantly, it seems that peak optical spectra may not necessarily reveal all explosions that harbor a magnetar. 
SN\,2020wnt shows us that the spectroscopic signatures related to a central engine may be hidden near peak in events with large ejecta mass and lower energy magnetars. 
Nebular spectroscopy of luminous SNe that have SNe~Ibc-like spectra at peak is needed to constrain the fraction of SESNe harboring a central engine. 
It could very well be possible that \textit{all} SESNe (or even H-rich CCSNe) are affected by the neutron star formed in core collapse to a different degree. 
Late-time observations, both photometric and spectroscopic, with near-IR coverage to measure a realistic bolometric luminosity, are needed for a large sample of these explosion to measure the distribution of the initial rotation and the magnetic field of a neutron star born out of the core collapse process.

\section*{Acknowledgement}
We thank Norbert Langer, Matt Nicholl, Conor Omand, Dan Kasen, David Khatami, Anders Jerkstrand for helpful discussions. %
We also thank the organizers of the IAU360 Massive Stars Near and Far symposium where a lot of helpful discussions about this object were made. 
We thank Robert Quimby for providing data points used to make Fig.~\ref{fig:type_score}.
We thank Matt Nicholl for providing the mean SLSN and SN Ic spectra for Fig.~\ref{fig:spec_comparison}.
S. Schulze acknowledges support from the G.R.E.A.T research environment, funded by {\em Vetenskapsr\aa det},  the Swedish Research Council, project number 2016-06012.
A.G. acknowledges support from the Flatiron Institute Center for Computational Astrophysics Pre-Doctoral Fellowship Program in Spring 2022. 
A.G. is also supported by the Illinois Distinguished Fellowship, the National Science Foundation Graduate Research Fellowship Program under Grant No. DGE – 1746047, and the Center for Astrophysical Surveys Graduate Fellowship at the University of Illinois.
Some of the data presented herein were obtained at the W. M. Keck Observatory, which is operated as a scientific partnership among the California Institute of Technology, the University of California and the National Aeronautics and Space Administration. The Observatory was made possible by the generous financial support of the W. M. Keck Foundation.
NIRES data presented in this paper were supported in part by NASA Keck PI Data Awards 2020B\_N141 and 2021A\_N147 (PI: Jha), administered by the NASA Exoplanet Science Institute. 
Parts of this work are based on observations obtained at the international Gemini Observatory, a program of NSF’s NOIRLab, which is managed by the Association of Universities for Research in Astronomy (AURA) under a cooperative agreement with the National Science Foundation on behalf of the Gemini Observatory partnership: the National Science Foundation (United States), National Research Council (Canada), Agencia Nacional de Investigaci\'{o}n y Desarrollo (Chile), Ministerio de Ciencia, Tecnolog\'{i}a e Innovaci\'{o}n (Argentina), Minist\'{e}rio da Ci\^{e}ncia, Tecnologia, Inova\c{c}\~{o}es e Comunica\c{c}\~{o}es (Brazil), and Korea Astronomy and Space Science Institute (Republic of Korea).
This work uses data from Infrared Telescope Facility, which is operated by the University of Hawaii under contract 80HQTR19D0030 with the National Aeronautics and Space Administration.
The authors wish to recognize and acknowledge the very significant cultural role and reverence that the summit of Maunakea has always had within the indigenous Hawaiian community.  We are most fortunate to have the opportunity to conduct observations from this mountain.
Parts of this work are based on observations obtained with the Samuel Oschin Telescope 48-inch and the 60-inch Telescope at the Palomar Observatory as part of the Zwicky Transient Facility project. ZTF is supported by the National Science Foundation under Grants No. AST-1440341 and AST-2034437 and a collaboration including current partners Caltech, IPAC, the Weizmann Institute of Science, the Oskar Klein Center at Stockholm University, the University of Maryland, Deutsches Elektronen-Synchrotron and Humboldt University, the TANGO Consortium of Taiwan, the University of Wisconsin at Milwaukee, Trinity College Dublin, Lawrence Livermore National Laboratories, IN2P3, University of Warwick, Ruhr University Bochum, Northwestern University and former partners the University of Washington, Los Alamos National Laboratories, and Lawrence Berkeley National Laboratories. Operations are conducted by COO, IPAC, and UW.
The ZTF forced-photometry service was funded under the Heising-Simons Foundation grant \#12540303 (PI: Graham).
Parts of this work are based on observations made with the Nordic Optical Telescope, owned in collaboration by the University of Turku and Aarhus University, and operated jointly by Aarhus University, the University of Turku and the University of Oslo, representing Denmark, Finland and Norway, the University of Iceland and Stockholm University at the Observatorio del Roque de los Muchachos, La Palma, Spain, of the Instituto de Astrofisica de Canarias.

\bibliography{2020wnt}{}
\bibliographystyle{aasjournal}

\newpage
\appendix 

\setcounter{table}{0}
\renewcommand{\thetable}{A\arabic{table}}

\setcounter{figure}{0}
\renewcommand{\thefigure}{A\arabic{figure}}

\begin{table}[!ht]
\caption{Log of spectroscopic observations}
\label{tab:spec_log}
\centering
\begin{tabular}{llll}
\toprule
Date (UT)       & MJD & Days from peak & Telescope/Instrument \\
\hline 
\multicolumn{4}{c}{\textit{Optical}} \\
\hline
2020-11-11                     & 59164                   & -47                       & P200/DBSP            \\
2020-11-15                     & 59168                   & -43                       & Shane/Kast           \\
2020-12-06                     & 59189                   & -23                       & NOT/ALFOSC           \\
2020-12-14                     & 59197                   & -15                       & NOT/ALFOSC           \\
2020-12-22                     & 59205                   & -7                        & Shane/Kast           \\
2021-01-06                     & 59220                   & 6                         & Shane/Kast           \\
2021-01-07                     & 59221                   & 7                         & P200/DBSP            \\
2021-01-11                     & 59225                   & 11                        & Shane/Kast           \\
2021-01-20                     & 59234                   & 20                        & NOT/ALFOSC           \\
2021-02-06                     & 59251                   & 36                        & Shane/Kast           \\
2021-02-11                     & 59256                   & 41                        & Shane/Kast           \\
2021-02-19                     & 59264                   & 49                        & Shane/Kast           \\
2021-03-22                     & 59295                   & 79                        & Shane/Kast           \\
2021-04-06                     & 59310                   & 93                        & Shane/Kast           \\
2021-07-15                     & 59410                   & 190                       & P200/DBSP            \\
2021-07-15                     & 59410                   & 190                       & Shane/Kast           \\
2021-07-28                     & 59423                   & 203                       & NOT/ALFOSC           \\
2021-08-03                     & 59429                   & 209                       & Shane/Kast           \\
2021-08-06                     & 59432                   & 212                       & Keck/LRIS            \\
2021-08-12                     & 59438                   & 217                       & Keck/LRIS            \\
2021-08-13                     & 59439                   & 218                       & Shane/Kast           \\
2021-11-06                     & 59524                   & 301                       & Keck/LRIS            \\
2022-01-05                     & 59584                   & 359                       & Keck/LRIS            \\
2022-09-22                     & 59844                   & 611                       & Keck/LRIS \\
\hline 
\multicolumn{4}{c}{\textit{Near-Infrared}} \\
\hline
2020-12-24                     & 59207                   & -5                        & Keck/NIRES           \\
2021-02-04                     & 59249                   & 34                        & P200/TSpec           \\
2021-02-22                     & 59267                   & 52                        & IRTF/SpeX            \\
2021-02-23                     & 59268                   & 53                        & Keck/NIRES           \\
2021-03-24                     & 59297                   & 81                        & Gemini/GNIRS         \\
2021-09-05                     & 59462                   & 241                       & IRTF/SpeX            \\
2021-11-19                     & 59537                   & 313                       & Gemini/GNIRS         \\
2022-01-25                     & 59604                   & 379                       & \textit{HST}/WFC3    \\
\hline
\end{tabular}
\end{table}
 
\begin{table}
\begin{threeparttable}
\caption{Emission-line measurements and derived spectral properties of the nucleus of the host galaxy of SN\,2020wnt}
\label{tab:hostgalaxyemissionlines}
\centering
\begin{tabular}{ccc}
\toprule 
Measurement & Unit/Wavelength & Value \\
\hline 
$A_V$ &~mag & 0.006$^{+0.173}_{-0.006}$\\
SFR H$\alpha$ & M$_\odot$~yr$^{-1}$ & 0.075$^{+0.014}_{-0.001}$ \\
T$_e$(\protect[\ion{O}{2}]) & K & $11991^{+388}_{-415}$  \\
T$_e$(\protect[\ion{O}{3}]) &  K & $12184^{+472}_{-471}$\\
12+log$_{10}$(O+/H+) & & $7.54^{+0.07}_{-0.06}$ \\
12+log$_{10}$(O++/H+) & & $7.91^{+0.06}_{-0.05}$ \\
12+log$_{10}$(O/H) & & $8.06\pm0.06$ \\
Z & Z$_\odot$ &$0.24\pm0.03$\\
H$\alpha$      & 6563 & 389.9$\pm$2.8   \\
H$\beta$       & 4861 & 136.1$\pm$5.7   \\
H$\gamma$      & 4340 & 62.3$\pm$3.2    \\
\protect[\ion{O}{2}]  & 3727 & 260.0$\pm$11.5  \\
\protect[\ion{O}{3}] & 4363 & 6.9$\pm$0.7     \\
\protect[\ion{O}{3}] & 4959 & 200.6$\pm$4.6   \\
\protect[\ion{O}{3}] & 5007 & 582.7$\pm$19.4  \\
\protect[\ion{N}{2}]  & 6548 & 5.8$\pm$1.8     \\
\protect[\ion{N}{2}]  & 6584 & 14.6$\pm$2.1    \\
\protect[\ion{S}{2}]  & 6716 & 36.3$\pm$0.4    \\
\protect[\ion{S}{2}]  & 6731 & 25.8$\pm$1.7    \\
\protect[\ion{O}{2}]  & 7320 & 3.5$\pm$0.5   \\
\protect[\ion{O}{2}]  & 7330 & 4.0$\pm$0.8   \\
\protect[\ion{S}{3}] & 9069 & 27.5$\pm$0.5    \\
\protect[\ion{S}{3}] & 9531 & 50.8$\pm$0.7    \\
\hline \\
\end{tabular}
\begin{tablenotes}
\item The Balmer decrement is calculated assuming the \citet{Calzetti2000} extinction law. The SFR is derived using the \citet{Kennicutt1998} relation. All emission lines fluxes are corrected for Galactic foreground extinction \citep{schlafly2011}, but are not corrected for the internal attenuation within the host system. Line flux values are in units of 10$^{-17}$ erg cm$^{-2}$ s$^{-1}$\AA$^{-1}$ and uncertainties are 1-$\sigma$.
\end{tablenotes}
\end{threeparttable}
\end{table}
\begin{table}
    \begin{threeparttable}
    \caption{Photometry of SN\,2020wnt's host galaxy}
    \label{tab:hostgalaxyphotometry}
    \centering
        \begin{tabular}{cccc}
        \hline
         Filter & Magnitude & Instrument & Reference \\
        \hline
$g$       &  19.35 $\pm$ 0.12 &  PS1       & This work \\
$r$       &  18.67 $\pm$ 0.10 &  PS1       & This work \\
$i$       &  18.38 $\pm$ 0.11 &  PS1       & This work \\
$z$       &  18.22 $\pm$ 0.11 &  PS1       & This work \\
$y$       &  18.05 $\pm$ 0.25 &  PS1       & This work \\
        \hline
        \end{tabular}
        \begin{tablenotes}
        \item 
        Magnitudes are not corrected for Galactic foreground extinction.
        \end{tablenotes}
    \end{threeparttable}
    \end{table}

\begin{figure}
\centering
\includegraphics[width=0.5\linewidth]{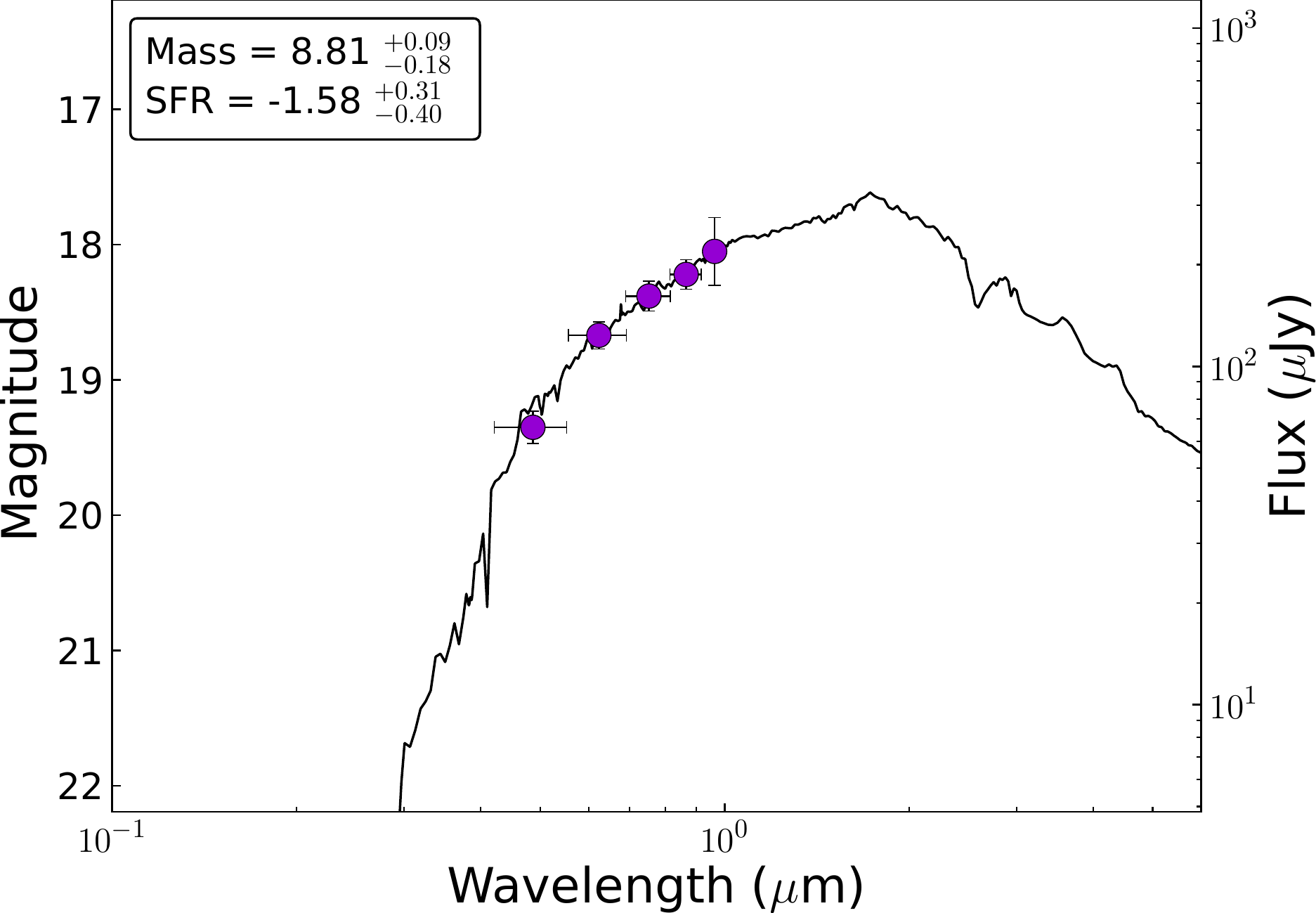}
    \caption{Fit to the spectral energy distribution of the host of 2020wnt using the SED modeling code \texttt{Le Phare} \citep{Ilbert2006}. Observed PS1 photometry is plotted in purple.}
    \label{fig:2020wnt_galaxy_sed}
\end{figure}

\begin{figure}
\includegraphics[width=0.49\textwidth]{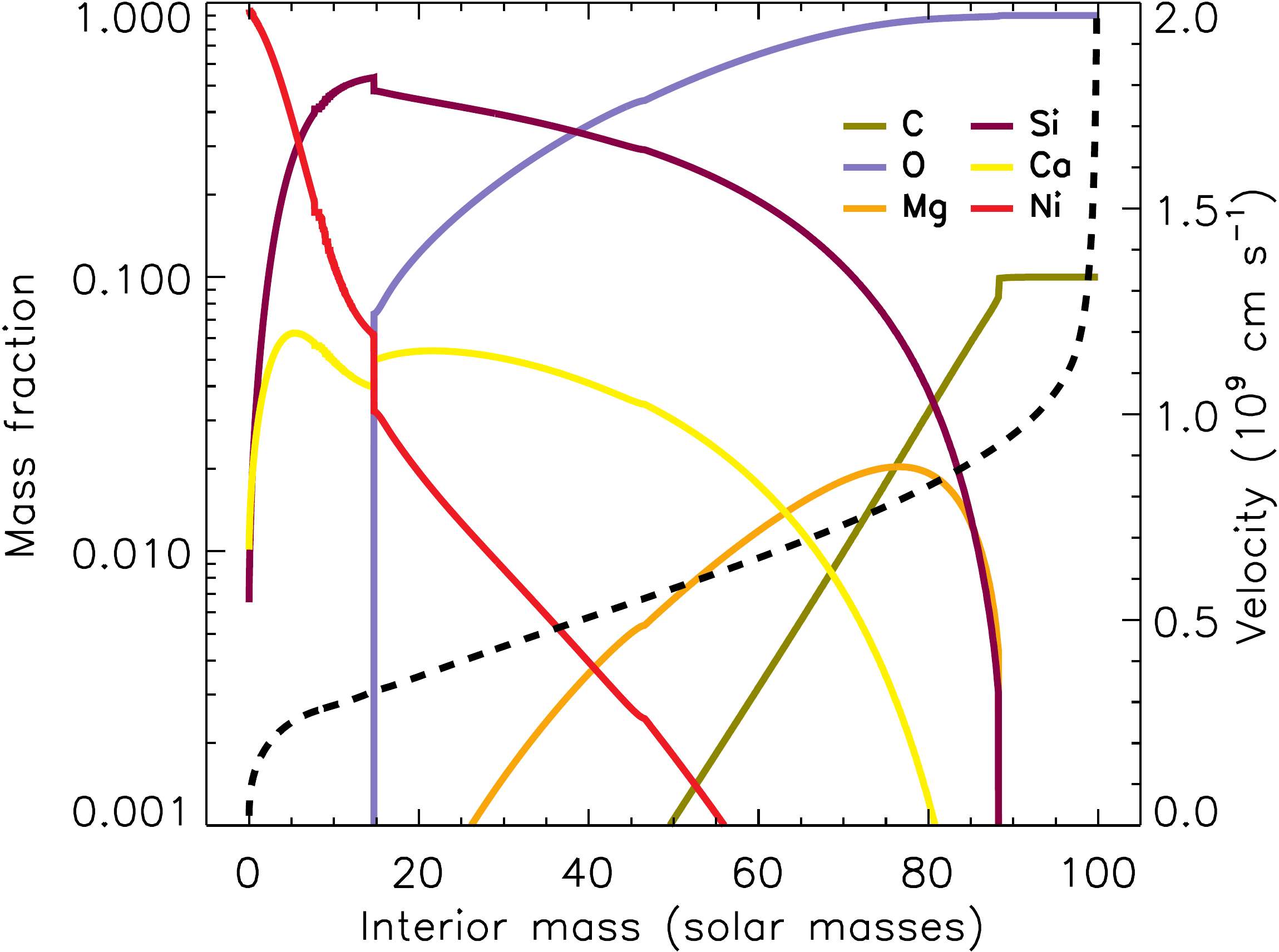} \hfill
\includegraphics[width=0.49\textwidth]{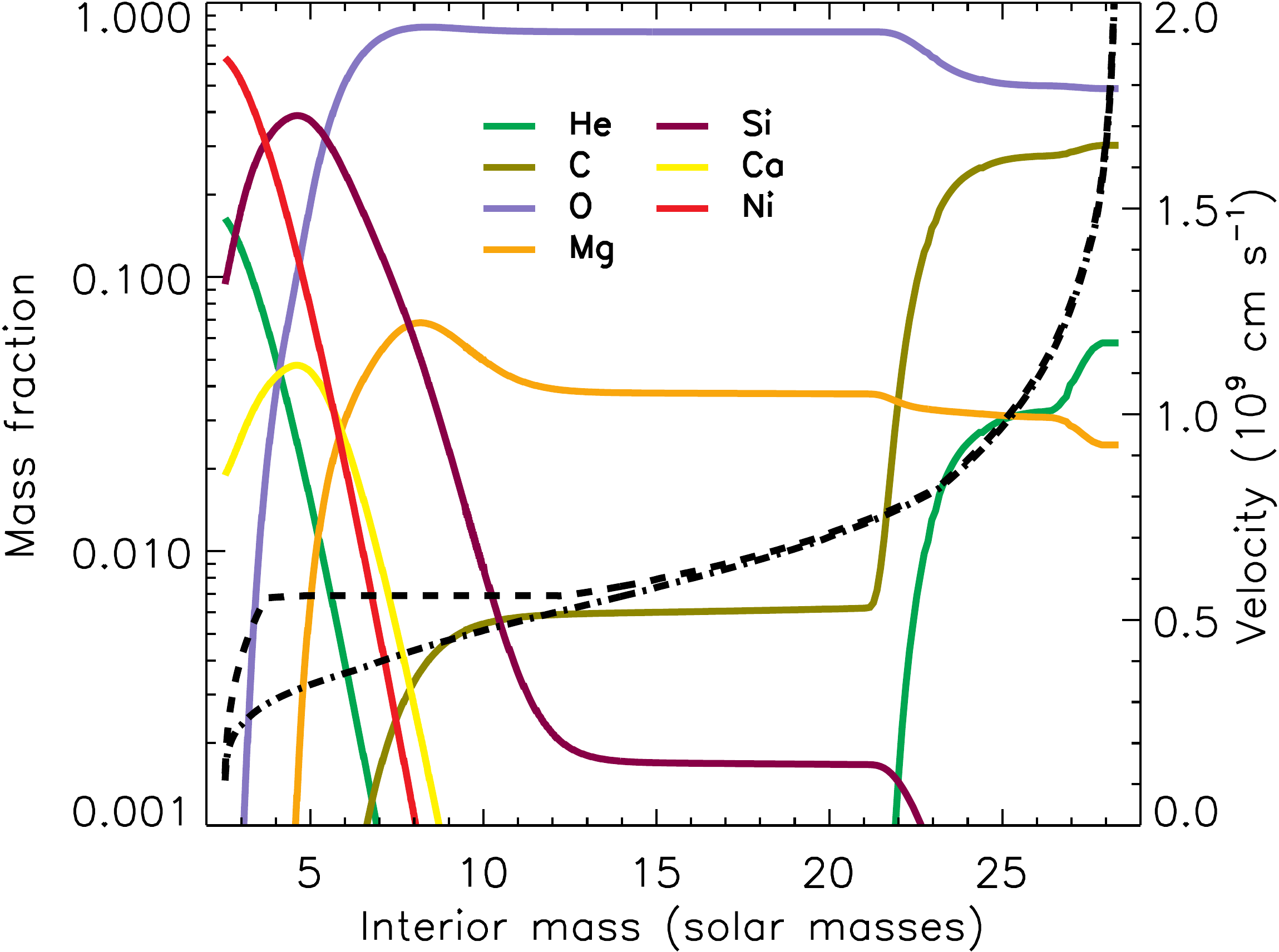}
\caption{\textbf{Left:} Velocity (dashed line) and composition of Model C100 (Table~\ref{tab:new_models}). The
  explosion energy was $4.37 \times 10^{52}$ erg and 5.13 \Msun \ of
  $^{56}$Ni was synthesized. These velocities are not dissimilar to SN\,2020wnt.
  \textbf{Right:} Composition and velocity of the Model T50 (Table~\ref{tab:new_models}). 
  The dot dashed line and the dashed line are the velocity profiles of the ejecta without and with the magnetar, respectively.
  Note the nearly constant velocity with the magnetar in the inner 11 \Msun, which is a consequence of the
  pulsar bubble sweeping up matter into a thin shell. A moderate amount of mixing has been artificially applied to the composition.} 
\label{fig:vel_composition}
\end{figure}

\end{document}